\PassOptionsToPackage{usenames,dvipsnames}{color}
\documentclass[%
floatfix,
showkeys,
 reprint,
superscriptaddress,
nofootinbib,
 amsmath,amssymb,
 amsfonts,
 aps,
prd]{revtex4-2}
\usepackage{graphicx}
\usepackage{enumitem}

\usepackage{tikz}
\usepackage{macros}
\usepackage{dcolumn}
\usepackage{bm}
\usepackage{hyperref}
\usepackage[caption=false]{subfig}
\usepackage{graphicx}
\hypersetup{
    colorlinks=true,
    linkcolor=blue,
    filecolor=blue,      
    urlcolor=blue,
    citecolor=blue, 
    pdftitle={PTA Global Gibbs Analysis},
    pdfpagemode=FullScreen,
    }
\usepackage{appendix}
\usepackage{makecell}

\usepackage{booktabs}
\usepackage[normalem]{ulem}
\usepackage{tabularx}
\newcolumntype{Y}{>{\centering\arraybackslash}X}
\usepackage[usenames,dvipsnames]{color}

\begin{document}

%\linenumbers

\preprint{APS/123-QED}

\title{Solving the PTA Data Analysis Problem with a Global Gibbs Scheme}
\author{Nima Laal}
\affiliation{Department of Physics and Astronomy, Vanderbilt University, 2301 Vanderbilt Place, Nashville, TN 37235, USA}
 \email{nima.laal@gmail.com}

 \author{Stephen R. Taylor}
\affiliation{Department of Physics and Astronomy, Vanderbilt University, 2301 Vanderbilt Place, Nashville, TN 37235, USA}

\author{Rutger van Haasteren}
\affiliation{Max-Planck-Institut f{\"u}r Gravitationsphysik (Albert-Einstein-Institut), Callinstra{\ss}e 38, D-30167, Hannover, Germany\\
Leibniz Universit{\"a}t Hannover, D-30167, Hannover, Germany}

 \author{William G Lamb}
\affiliation{Department of Physics and Astronomy, Vanderbilt University, 2301 Vanderbilt Place, Nashville, TN 37235, USA}

\author{Xavier Siemens}
\affiliation{Department of Physics, Oregon State University, 1500 SW Jefferson Way, Corvallis, OR 97331, USA}

\date{\today}

\begin{abstract}
The announcement in the summer of 2023 about the discovery of evidence for a gravitational wave background (GWB) using pulsar timing arrays (PTAs) has ignited both the PTA and the larger scientific community's interest in the experiment and the scientific implications of its findings. As a result, numerous scientific works have been published analyzing and further developing various aspects of the experiment, from performing tests of gravity to improving the efficiency of the current data analysis techniques. In this regard, we contribute to the recent advancements in the field of PTAs by presenting the most general, agnostic, per-frequency Bayesian search for a low-frequency (red) noise process in these data. Our new method involves the use of a conjugate Jeffrey's-like multivariate prior which allows one to model all unique parameters of the global PTA-level red noise covariance matrix as a separate model parameter for which a marginalized posterior-probability distribution can be found using Gibbs sampling. Even though perfecting the implementation of the Gibbs sampling and mitigating the numerical stability challenges require further development, we show the power of this new method by analyzing realistic and theoretical PTA simulated data sets. We show how our technique is consistent with the more restricted standard techniques in recovering both the auto and the cross-spectrum of pulsars' low-frequency (red) noise. Furthermore, we highlight ways to approximately characterise a GWB (both its auto- and cross-spectrum) using Fourier coefficient estimates from single-pulsar and so-called CURN (common uncorrelated red noise) analyses via analytic draws from a specific Inverse-Wishart distribution.   
\end{abstract}


\maketitle

\section{\label{sec:Introduction}Introduction}
Pulsar timing arrays (PTA) \citep{Detw, Shazin, 1990ApJ...361..300F, taylor2021nanohertz} are galactic-scale experiments aimed at exploring the universe through the recording of the times of arrival (TOA) of radio-frequency pulses as they sweep by the line-of-sight of radio telescopes. If observed persistently and for a sufficiently long time, the TOA data collected will eventually enable one to detect subtle deviations from the expected deterministic behaviour of radio signals as they arrive at earth, unlocking the secrets of the astrophysical or cosmological processes responsible for the detected deviations. Of particular interest are the gravitational-wave-induced deviations whose detection has been the primary goal of the international pulsar-timing community for at least two decades.

After decades of observing millisecond-period pulsars, the International Pulsar Timing Array (IPTA) found evidence for a gravitational-wave background (GWB) \citep{15yr, d2, d3}. The findings leading to the announcement in the summer of 2023 came after years of work by the constituents of the IPTA analyzing different and independent PTA data sets, with the North American Nanohertz Observatory for Gravitational Waves (NANOGrav) reporting the highest statistical significance for a GWB signal across its 67 analyzed millisecond-period pulsars. The findings of European Pulsar Timing Array (EPTA), Indian pulsar timing array (InPTA), and Parkes pulsar timing array (PPTA) uncover compatible and complementary evidence for a GWB, albeit with varying degrees of statistical significance \citep{IPTACOMP}. 

As predicted theoretically in \citet{CURNFIRST} and through simulations in \citet{Astroforcast}, the evidence for a GWB first appeared in the form of a common but uncorrelated low-frequency noise process several years before the recent announcements by IPTA. The seminal works of \citet{12year} and \citet{DR2} detected such a process and presented a GWB data analysis pipeline that became the foundation for the future GWB data analysis efforts, culminating in the most recent IPTA results. By the standards agreed upon by the constituents of the IPTA \citep{checklist}, what constitutes evidence for the GWB is not a single number, nor is it about the results of a single analysis: the correlation signature induced by the GWB on the timing data of many pairs of pulsars needs to be observed using a slew of rigorously tested Bayesian and frequentist statistical techniques whose limitations and strengths have been well-established prior to their application on the real data set. The confidence in the findings of the IPTA stems from upholding such standards. 

Yet, despite the uncovered evidence for a GWB and its consistency with astrophysical expectations \citep{15yrastro}, as well as selected cosmological models \citep{15yrnewphysics}, characterization of the GWB using PTAs is a very active area of research; the complete spectral features as well as the polarization properties of the GWB are not yet fully determined \citep{15yealtpol}. Even though increasing sensitivity from future data releases will undoubtedly help in answering the remaining questions, there is still a need for robust GW data analysis techniques to better probe the GWB and help with finalizing its characterization. 

In particular, a truly model-agnostic (or minimally modeled) per-frequency estimation of the low-frequency correlated and uncorrelated noise of each pulsar is not within the capability of the current GWB detection techniques.  Since the signature of the inter-pulsar correlations induced by the GWs from some modified theories of gravity are predicted to vary significantly with GW frequency \citep{Lee_2008, sublu}, an agnostic per-frequency Bayesian pipeline could be invaluable to tests of beyond-GR gravity efforts using PTAs. %Furthermore, as shown in \citet{Kaiser_2022}, resolving source confusion (i.e., disentangling multiple astrophysical and cosmological sources for a GWB) is an extremely difficult task even when attempted on simulated PTA data sets. Perhaps a model-agnostic per-frequency data analysis technique could provide a better chance of succeeding at such attempts. 

Additionally, there is an ongoing need to build a robust pipeline to estimate various global PTA-level quantities (e.g., the GWB correlation signature and its power spectral density) from the results of single-pulsar analyses. This need is based on the fact that many pulsars are affected by chromatic (i.e., radio-frequency dependent) noise whose properties are unique to (the line-of-sight path to) individual pulsars and require bespoke noise modeling.
The inclusion of such customized models for every affected pulsar in the global PTA-level analysis is practically impossible at present \citep{meerkat, 15yrANM, Null}. Hence, there is a need for techniques capable of achieving this goal.

It is guided by this overarching aim that we contribute to the fast-expanding literature of PTA GW searches by introducing the most agnostic, general, and fully Bayesian search for a low-frequency (red) noise Gaussian process to date. Our technique is an expansion of our previous work in \citet{Laal:2023etp}, where we explored the capabilities of Gibbs sampling for single pulsar analyses. In this work, we expand the technique to include more pulsars. That is, we derive and further explore a multi-pulsar Gibbs sampling routine whose underlying data analysis approach is agnostic and model-independent: every single unique entry in the global PTA-level red noise covariance matrix is treated as a model parameter. The only assumptions we make about the red noise process are $(1)$ its stationarity (i.e., frequency bins are independent), and $(2)$ its Gaussianity.

What our new technique offers is different from other recent advancements in the field of GW and GWB data analysis techniques for PTAs \citep{parallel, fitting, 2024PhRvL.133a1402S, Rutger1, GF, discovery, OSperFreq}. Although fast, the novelty of the technique is not in its speed but rather in its generality. This method is capable of searching for both auto- and cross- correlation content of a PTA's red noise process without being bounded by a specific parametrization or assumptions of noise models. Most notably, our new technique allows for different frequency bins to have independent correlation structures, a feature lacking in the current GWB data analysis techniques. Knowing the dependence of the GWB statistics to noise modeling \citep{Hazboun_2020}, our new method can provide fresh insights on the recent and future discoveries of the IPTA. Lastly, one can use the results of single pulsar analyses, namely Fourier coefficients, to approximate both the correlation and the power spectral density of the GWB using a modified version of our data analysis method. This feature could prove useful to \emph{advanced noise modeling} efforts \citep{15yrANM} for determining the impact of customized noise models of pulsars on the characterization of the GWB.

The paper is structured as follows. In \S\ref{sec:Methods}, we review the PTA noise modeling structure by defining relevant quantities and introducing the most common noise modeling assumptions. In \S\ref{sec:Bayesian Modeling}, we perform a detailed derivation and further discuss the new multi-pulsar Gibbs sampling method. In \S\ref{sec:Gibbs}, the implementation challenges and the flow of the Gibbs sampling is presented. In \S\ref{sec:Resamp}, we introduce the basics of the post-processing tools we utilize to process the output of Gibbs sampling. In \S\ref{sec:sims}, we explore the capabilities of the presented multi-pulsar Gibbs sampling using realistic and theoretical PTA simulated data sets. In \S\ref{sec:IWdraws}, we present how the result of single-pulsar analysis can be used to approximately characterize an underlying GWB. Finally, in \S\ref{sec:conclusion}, we present our concluding remarks.

\section{\label{sec:Methods}The PTA Data Model}

\subsection{Single pulsar}

In pulsar timing analysis, the result of fitting a deterministic timing ephemeris model to a set of TOAs and subsequently subtracting the best-fit solution from these TOAs, is a set of timing residuals for each pulsar, $\bm{\delta t}$. The model with which we analyze these timing residuals is 
\begin{align} \label{firsteq}
    \bm{\delta t} =M \bm{\epsilon} +F\bm{a}+\bm{n},
\end{align}
where the first term accounts for deviations from the best-fit timing solution, such that $M$ is the timing-model design matrix whose elements are partial derivatives of the TOAs with respect to each timing model parameter, while $\bm{\epsilon}$ is a vector of deviation parameters in our linearized treatment of the timing model. The second term describes all low-frequency processes as a Fourier series, such that $F$ is a Fourier design matrix with dimensions $n_\mathrm{TOA} \times 2n_f$, containing alternating length-$n_\mathrm{toa}$ columns of sine and cosine evaluations at each sampling frequency of the pulse time series, and $\bm{a}$ are Fourier coefficients of the low-frequency process. Finally, $\bm{n}$ is a time series of white noise in the TOAs that can be ascribed to various forms of measurement uncertainty. For practical reasons and compactness we can combine the first two terms in \autoref{firsteq} into $T\bm{b} = M \bm{\epsilon} +F\bm{a}$, where $T = [M\quad F]$ and $\bm{b}$ is the concatenation of $\bm{\epsilon}$ and $\bm{a}$. 

By assuming a Gaussian likelihood for white noise in the TOAs, one can write
\begin{equation}
p(\bm{\delta t} | \bm{b}) = \mathcal{N}(T\bm{b}, N)
\end{equation}
such that the mean is $T\bm{b}$, and the white noise covariance $N$ in each pulsar is composed of squared TOA uncertainties and other calibrating factors (e.g., EFAC, EQUAD, ECORR defined in \citet{9Year}) that are determined prior to any global PTA analysis.\footnote{This procedure of fixing parameters in a global PTA analysis that have previously been searched for at the single-pulsar level can be justified by the fact that these white noise parameters are typically not highly covariant with other processes.} The parameters $\bm{b}$ are typically not of primary interest as their covariance across many realizations of the noise is the quantity that can reveal information about the power at low-frequencies in each pulsar as well as in inter-pulsar correlated processes like the GWB. Hence, we introduce $B$ as the covariance of $\bm{b}$, and use Bayes' Theorem to write
\begin{equation} \label{bayes}
    p(\bm{b}, B | \bm{\delta t}) \propto p(\bm{\delta t} |\bm{b}) \times p( \bm{b}|B) \times p(B).
\end{equation}
We now make the additional assumption that our low-frequency processes and timing model offsets can be described as random Gaussian processes with $B$ as the covariance matrix, such that $p(\bm{b}| B) = \mathcal{N}(\bm{0}, B)$ and 
\begin{align} \label{BINV}
       \left\langle \bm{b}\bm{b}^T \right\rangle = B &=\left[ \begin{matrix}
   \bm{\infty} & 0  \\
   0 & \Phi   \\
\end{matrix} \right],
\end{align}
with $\Phi$ as the covariance matrix of dimensions $2n_f\times 2n_f$ on the Fourier coefficients $\bm{a}$. Notice that the $\bm{\infty}$ simply means that we have placed an unbounded improper prior on the linear timing model offsets; this presents no practical issues since the likelihood is strongly peaked in these parameters, and furthermore we are typically only interested in the inverse covariance matrix, such that $B^{-1} = \mathrm{diag}(0, \Phi^{-1})$. This is a common practice in the PTA detection efforts \citep{Lentati:2012xb, Gibbs0, LowRankPaper}. In software implementations, the large numerical value of $10^{40}$ is frequently used instead of $\bm{\infty}$ for practical reasons.

\subsection{Multi-pulsar} \label{sec:multipulsar_datamodel}

In going from single-pulsar analyses to a model for the multi-pulsar analysis, we note that the multi-pulsar likelihood $p(\{\delta t\} | \{\bm{b}\})$ is simply a product over pulsars, such that
\begin{equation} \label{multipulsar_like}
    p(\{\delta t\} | \{\bm{b}\}) = \prod\limits_{I=1}^{n_p} p(\delta t_I | \bm{b}_I),
\end{equation}
where $\{\cdot\}$ denotes the set of quantities for all pulsars, and $I$ indexes over $n_p$ pulsars.

In the following it will be convenient to suppress the $\{\cdot\}$ notation, which means that we will redefine some quantities from the previous single-pulsar data-model description. $M$ and $F$ become block matrices over pulsars in which the individual pulsar timing-model design matrices and Fourier design matrices are inserted, respectively, i.e.,
\begin{equation}
M = \left[\begin{matrix}
          M_1 & 0  & \hdots & 0 \\
          0 & M_2 & \hdots & 0\\
          \vdots & \vdots & \ddots &  \vdots \\
          0 & 0 & \hdots & M_{n_p}\\
    \end{matrix} \right] 
\end{equation}
and similar for $F$. The PTA $T$ matrix that groups these together, and the PTA block-diagonal white noise covariance matrix $N$, follow a similar pattern. 

The PTA covariance matrix $B$ becomes a block matrix of dimension $2n_p n_f \times 2n_p n_f$, such that
\begin{equation}
B = \left[\begin{matrix}
          B_{11} & B_{12}  & \hdots & B_{1 n_p} \\
          B_{21} & B_{22} & \hdots & B_{2 n_p}\\
          \vdots & \vdots & \ddots &  \vdots \\
          B_{n_p 1} & B_{n_p 2} & \hdots & B_{n_p n_p}\\
    \end{matrix} \right],
\end{equation}
and finally the vectors $\bm{\delta t}$ and $\bm{b}$ can be considered as concatenated quantities over the PTA.

With these updated definitions for various vector and matrix quantities, as well as \autoref{multipulsar_like}, \autoref{bayes} now describes the multi-pulsar PTA posterior on $\bm{b}$ and its covariance $B$. In fact, the portion of $B$ that is data dependent is really just $\Phi$. The key issue then becomes an appropriate choice of prior on $\Phi$, the covariance matrix of the low-frequency processes in the data.

\section{\label{sec:Bayesian Modeling}Conjugate priors \& conditional posteriors}

Often in PTA analyses, $\Phi$ is assumed to take a parameterized functional form. For example, for the covariance between pulsars $I$ and $J$,
\begin{equation} \label{eq:plaw_irn_hd}
    \Phi_{kk';IJ} = \delta_{IJ}\delta_{kk'}\kappa_{k;I} + \Gamma_{IJ}\delta_{kk'}\rho_{k},
\end{equation}
where $\kappa_{k;I}$ describes intrinsic red noise\footnote{In the context of this work, intrinsic red noise is any noise that is not GWB; hence, it is unique to a given pulsar.} in pulsar $I$ that is often assumed to be a power-law over frequencies (indexed by $k$), and $\Gamma_{IJ}$ is the correlation signature (or ``overlap reduction function'' for a GWB) of an inter-pulsar correlated process with a power-law form $\rho_k$ over frequencies.\footnote{The Kronecker deltas $\delta_{IJ}$ and $\delta_{kk'}$ describe independence between pulsars and frequencies, respectively. Note that we will continue to use the notation throughout of a semicolon separating frequency and pulsar indices.} With a power-law functional form, we have two parameters to determine for each process, and each of these parameters has its own prior. In conventional acceptance-rejection Metropolis-Hastings Markov chain Monte Carlo (MCMC) sampling of the posterior probability distribution, this would be the last decision we need to make about our model; we have priors on the parameters of power-law (or other) descriptions of low-frequency processes in our data, and the effort then shifts to efficient sampling of the parameter and hyperparameter space. The nature of MCMC rejection-based sampling makes it less efficient as the number of parameters increases.

However there is a special class of priors that allows one to avoid acceptance-rejection sampling. By identifying priors on model parameters that result in \textit{conditional} posteriors on parameters belonging to the same family of (standard) distributions as the prior, one can directly draw samples with large step sizes, with no rejection necessary. These are called \textit{conjugate priors}, which we describe in more detail next, and are key to our new multi-pulsar Gibbs scheme.

\subsection{\label{sec:prior}Prior probability}

In \citet{Gibbs0} and \citet{Laal:2023etp}, it was shown how, in a single-pulsar analysis, the following choice of a bounded log-uniform prior on the diagonal elements of $\Phi$,
\begin{align}
    \begin{split}
        p(\Phi_{II}) &= \prod\limits_{k=1}^{n_f}\frac{1}{\Phi_{k;II}},\\
        \log \left( {{\Phi_{k;II} }} \right)_{\min }&<\log \left( {{\Phi }_{k;II}} \right)<\log \left( {{\Phi_{k;II} }} \right)_{\max }\label{sgp},
    \end{split}
\end{align}
leads to a truncated Inverse-Gamma ($\text{IG}$) distribution (with shape parameter $\alpha$, and scale parameter $\beta$) for the conditional posterior probability of $p\left( \left. {{\Phi }_{II}} \right|{\bm{a}_{I}} \right)$ in each frequency bin, i.e.,
\begin{equation} \label{IG}
    p\left( \left. {{\Phi }_{II}} \right|{{\bm{a}}_{I}} \right)=\prod\limits_{k=1}^{{{n}_{f}}}{\text{IG}\left( \alpha =1,\beta =\frac{1}{2} \left[( a_{k;I}^c )^{2} + ( a_{k;I}^s )^{2}\right] \right)}.
\end{equation}
where $a_{k;I}^c$ and $a_{k;I}^s$ are the cosine and sine Fourier coefficients at frequency $k$ for pulsar $I$.\footnote{Implicit also is the model assumption of independence between frequency-bins.}

In the more general scenario where $\Phi$ describes the global PTA-level covariance matrix such that one must model inter-pulsar correlations, the above prior lacks the degrees of freedom necessary to describe all processes. We therefore need a new and more general prior. Upon maintaining the assumption about the independence of frequency-bins from each other, the family of conjugate priors for the covariance matrix $\Phi$ of a multivariate normal distribution can be written as \citep{gelmanbook}
\begin{align}
    p\left( \Phi  \right)=\prod\limits_{k=1}^{{{n}_{f}}}{{\left| \Phi_k  \right|}^{-\frac{{{\nu }}+{{n}_{p}}-1}{2}}}\text{exp}\left\{ -\frac{1}{2}\text{tr}\left( {{\kappa}_{0}}{{\Phi_k }^{-1}}\Psi_k  \right) \right\},
\end{align}
where $\Phi_k$ is the $n_p\times n_p$ covariance matrix at frequency $k$, $|\cdot|$ denotes the determinant, $\nu$ and $\kappa_0$ are constants, and $\Psi_k$ is another (arbitrary) $n_p \times n_p$ matrix. In order for this prior to be invariant under the scaling transformation $\Phi \to c\Phi $ for some nonzero constant $c$, one must take the limit as $\kappa_0$ approaches zero. This results in
\begin{align}
    p\left( \Phi  \right)=\prod\limits_{k=1}^{{{n}_{f}}}{{\left| \Phi_k  \right|}^{-\frac{{{\nu }}+{{n}_{p}}-1}{2}}} \label{general prior}.
\end{align}
We note several important points here: $(1)$ this prior applies to the matrix as a whole, rather than to its individual elements independently; $(2)$ the log-uniform prior for single-pulsar analyses in \autoref{sgp} is a special case of this prior when $n_p=1$ and $\nu = 2$; and $(3)$ even though the above prior guarantees invariance under a scaling transformation, it does not guarantee invariance under a logarithmic transformation of $\Phi_{II}$ parameters as \autoref{sgp} does. Philosophically, we intend for our new prior to behave similarly to a Jeffreys prior \footnote{Jeffreys prior is a popular choice for a non-informative prior derived form the mathematical structure of the likelihood function.}, in that it should not prefer one parameterization over another. Additionally, the prior should introduce minimal information to the model relative to the data. To achieve this behavior, careful selection of the parameter $\nu$ is crucial. We discuss the details of construction of the prior in the next section.

\subsection{\label{a}Conditional posterior probability}

\subsubsection{Coefficients %\texorpdfstring{$p(\bm{b} |\bm{B},\bm{\delta t})$}{p(Phi | a, delta t)}
}

The conditional posterior on the concatenated coefficients $\bm{b}$ given PTA covariance matrix $B$ is a straightforward generalization of what has already been described in \citet{Laal:2023etp} and \citet{Gibbs0}. Specifically, as discussed earlier in the \S\ref{sec:multipulsar_datamodel}, the $B$ matrix now contains off-diagonal blocks due to inter-pulsar correlated $\Phi$ matrices, i.e.,
\begin{equation} \label{bgivenrho}
    p(\bm{b} |B, \bm{\delta t}) = \mathcal{N}(\bm{\mu}, \Sigma), 
\end{equation}
where
\begin{align}
 \Sigma^{-1} &= T^T N^{-1} T + B^{-1} \label{sigma_Def}, \\ 
 \bm{\mu} &= \Sigma T^T N^{-1} \bm{\delta t} \label{mudef}.
\end{align}

\subsubsection{Covariance matrix%\texorpdfstring{$p(\Phi |\bm{a},\bm{\delta t})$}{p(Phi | a, delta t)}
}

The posterior probability for $\Phi$ conditioned on $\bm{a}$, $p(\Phi | \bm{a})$, can be found by normalizing $p\left( \left. \bm{b}, B \right|\bm{\delta t} \right)$ of  \autoref{bayes} with respect to $\Phi$ (i.e., $p\left( \left. \bm{b}, B \right|\bm{\delta t} \right)/p\left( \left. \bm{\delta t} \right| \bm{b}\right)$), and using the general conjugate prior of \autoref{general prior}. We then denote the outer-product between the coefficients $\bm{a}$ at frequency $k$ by an $n_p \times n_p$ symmetric matrix $S_k = S_k^{c} + S_k^{s}$, where
\begin{equation} \label{scalemat}
    S^c_{k;IJ} = a^c_{k;I} a^c_{k;J}, \quad S^s_{k;IJ} = a^s_{k;I} a^s_{k;J}. \\
\end{equation}
Upon assuming independence between frequency-bins one finds
\begin{align} \label{big-wishart}
    \begin{split}
       p\left( \left. \Phi  \right|\bm{a} \right) &= p\left( {{\Phi }} \right) \left[ \prod\limits_{k=1}^{n_f} \left| \Phi _k \right|^{-\frac{1}{2}} \exp \left\{ -\frac{1}{2}\text{tr}( S^c_k\Phi_k^{-1} ) \right\} \right. \\
       &\qquad\quad\left. \times \prod\limits_{k=1}^{n_f} \left| \Phi _k \right|^{-\frac{1}{2}} \exp \left\{ -\frac{1}{2}\text{tr}( S^s_k\Phi_k^{-1} ) \right\} \right] \\ 
     & =\prod\limits_{k=1}^{{{n}_{f}}}{{{\left| {{\Phi }_{k}} \right|}^{- \frac{\nu+n_p-1}{2} }}{{\left| {{\Phi }_{k}} \right|}^{-1}}\exp \left\{ -\frac{1}{2}\text{tr}\left[ {{S}_{k}}{{\Phi }_{k}}^{-1} \right] \right\}} \\ 
     & =\prod\limits_{k=1}^{{{n}_{f}}}{{{\left| {{\Phi }_{k}} \right|}^{-\frac{\nu +{{n}_{p}}+1}{2}}}\exp \left\{ -\frac{1}{2}\text{tr}\left[ {{S}_{k}}{{\Phi }_{k}}^{-1} \right] \right\}}\\
     &= \prod\limits_{k=1}^{{{n}_{f}}} \text{Inverse-Wishart} \left(\nu, {{S}_{k}}\right),
     \end{split}
\end{align}
where $\nu$ and $S_k$ can be identified as the \emph{degrees of freedom} and the \emph{scale-matrix}, respectively, of an Inverse-Wishart distribution \citep{gelmanbook}.

What remains to do is to choose a $\nu$ that results in a posterior probability that generalizes \autoref{IG}: the resulting conditional posterior probability for the diagonal elements $\Phi_{II}$ must match those of the corresponding Inverse-Gamma distribution regardless of the number of pulsars in the array. Subsequently, the only viable choice for $\nu$ is $\nu=n_p+1$ (see Appendix~\ref{sec:math proof} for a proof). Thus, we write\footnote{The form of the prior in \autoref{actual prior} was first introduced in \citet{GCprior} to study posterior distributions for multivariate normal parameters.}
\begin{align}
    p\left( \Phi  \right)&=\prod\limits_{k=1}^{{{n}_{f}}}{{\left| \Phi_k  \right|}^{-n_p}} \label{actual prior}, \\
    p\left( \left. \Phi  \right|\bm{a} \right)&= \prod\limits_{k=1}^{{{n}_{f}}} \text{Inverse-Wishart} \left(n_p+1, {{S}_{k}}\right) \label{rhogivenb}.
\end{align}

We note several interesting features about this choice of prior and the resulting conditional posterior. First, as we prove in Appendix~\ref{sec:math proof}, the Jeffrey's prior of \autoref{sgp} and our new choice of prior in \autoref{actual prior} share a common feature: the enforcement of the condition of invariance under logarithmic transformation of $\Phi_{II}$. Secondly, for $n_p = 1$ the conditional posterior probability distribution at each frequency corresponds to an Inverse-Wishart distribution with degrees of freedom $\nu = 2$ and $1 \times 1$ scale-matrix $S_k = ( a_{k;I}^c)^2 + (a_{k;I}^s)^2$ which is identical to \autoref{IG}. This is to be expected, as the Inverse-Wishart distribution is a generalization of the Inverse-Gamma distribution, and our choice of prior was informed by the requirement of obtaining the same conditional distribution on $\Phi_{II}$ as \autoref{IG} for for any $n_p$. Finally, this prior is improper. For practical reasons we must therefore impose lower and upper bounds on the $\Phi_{II}$ parameters, which are enforced during the numerical implementation of \autoref{actual prior}; this will be discussed later in \S\ref{sec:Gibbs}.

\subsection{\label{sec:Two-pulsar}Example: a PTA of two pulsars}

A two-pulsar PTA is the simplest non-trivial case to examine with our new prior and the resulting conditional posterior on $\Phi$. Without loss of generality, we can consider only one frequency and omit the index $k$ from the ensuing discussion. Based on \autoref{scalemat}, the scale matrix for the case of $n_p=2$ is
\begin{align}
    S=\left( \begin{matrix}
   \left( a_1^c \right)^2+\left( a_1^s \right)^2 & a_1^c a_2^c + a_1^s a_2^s  \\
   a_1^c a_2^c + a_1^s a_2^s & \left( a_2^c \right)^2 +\left( a_2^s \right)^2
\end{matrix} \right).
\end{align}
As we discuss in Appendix~\ref{sec:Sampling from an Inverse-Wishart Distribution}, the most computationally efficient method of sampling from an Inverse-Wishart distribution is the \emph{Bartlett Decomposition} \citep{bartlettmore,Bartlett_1934}. Upon applying this technique to our two-pulsar PTA, we find
\begin{align}
    Y_{11} \sim p(\Phi_{11}|\bm{a}_1);\quad Y_{11} &= \frac{ (a_1^c)^2 + (a_1^s)^2 }{X_{11,2}} \label{phi11}, \\ 
    Y_{22} \sim p(\Phi_{22}|\bm{a}_2);\quad Y_{22} &= \frac{ (a_2^c)^2 + (a_2^s)^2 }{X_{22,2}} \label{phi22},
\end{align}
where $X_{II,2}$ are draws from a $\chi^2$-distribution with two degrees of freedom. Since an Inverse-Chi-squared distribution is a special case of an Inverse-Gamma distribution, \autoref{phi11}, \autoref{phi22}, and \autoref{IG} (with one frequency) are all the same distributions. Therefore, as expected, the $\Phi_{II}$ parameters follow the same conditional posterior distributions as one would find in the case of $n_p=1$, which is a direct consequence of our prior choice.

For the cross-spectrum between pulsars $1$ and $2$ we have
\begin{align}
    Y_{12} &\sim p(\Phi_{12}|\bm{a}_1, \bm{a}_2), \\
    Y_{12} &= \frac{1}{X_{12,2}}\left( a_1^c a_2^c + a_1^s a_2^s -\frac{\left\| a_1^c a_2^s -a_1^s a_2^c \right\|}{\sqrt{X_{12,3}} / Z} \right) \label{phi12},
\end{align}
where $\Phi_{12} = \Phi_{21}$, $Z$ is drawn from a standard normal distribution, $X_{12,3}$ is drawn from a $\chi^2$-distribution with three degrees of freedom, $X_{12,2}$ is drawn from a $\chi^2$-distribution with two degrees of freedom, and $\| \cdot \|$ denotes an absolute value. Note that the five distributions, $X_{11,2}$, $X_{22,2}$, $X_{12,2}$, $X_{12,3}$, and $Z$, are not all independent from each other. For an $n_p \times n_p$ covariance matrix, there could exist only $n_p (n_p+1)/2$ independent entries in the matrix. Thus, in the case of $n_p=2$, only $3$ entries can be independent from each other. This further requires two out of $X_{11,2}$, $X_{22,2}$, $X_{12,2}$, $X_{12,3}$, and $Z$ to depend on others. In Bartlett decomposition, one makes the arbitrary choice that $X_{11,2}=X_{12,3}-Z^2$ and $X_{12,2}=X_{22,2}$\footnote{By properties of $\chi^2$ distributions, $\chi^2_{\nu+1} - Z^2=\chi^2_{\nu}$ for any $\nu$ larger than 1.}.
Despite the complicated shape of \autoref{phi12}, it remains the same regardless of the number of pulsars in the array; we demonstrate this numerically in Appendix~\ref{sec:math proof}. For a pair of pulsars with timing residuals containing low-frequency processes that are GWB-dominated, the term inside the absolute-value operator is nearly zero, making $\Phi_{12}$ follow an Inverse-Chi-squared distribution akin to $\Phi_{11}$ and $\Phi_{22}$.

\section{\label{sec:Gibbs}A multi-pulsar Gibbs scheme}

Gibbs sampling \citep{gibbsintro} is a class of MCMC algorithm in which parameters are updated sequentially through draws from their posteriors conditioned on the fixed values of other parameters. Once each parameter is updated, its value is held fixed in the conditional posterior of the next parameter's update, and so forth. Despite the fact that random samples are drawn from conditional probabilities of each parameter, over many iterations, the distribution of samples converges to the joint marginal posterior probability of all parameters as guaranteed by the principle of \emph{detailed balance}. In the field of PTAs, Gibbs sampling has been used previously in \citet{Gibbs0}, \citet{Laal:2023etp} and \citet{moreGibbs}.  

Usually conditional posteriors do not have distributions that belong to standard families from which we can directly draw samples, requiring samples to be proposed and assessed via the Metropolis-Hastings ratio to determine acceptance or rejection in the Markov chain. However, a special case of Gibbs sampling is when the conditional posterior probability distribution for all model parameters are used as proposal distributions within the MCMC. As a result, the acceptance rate of proposed samples is always $100\%$, as shown in Appendix~\ref{sec:GibbsMCMC}. The ability to draw random samples directly from conditional posterior probability distributions of parameters is made possible through the careful application of conjugate priors, as we have discussed earlier. In models with high parameter dimensionality, Gibbs sampling (especially with conjugate priors) provides an efficient and practical way to sample from the joint parameter posterior distribution.   

\subsection{Overview of the algorithm}
Here we outline a step-by-step implementation of Gibbs sampling for the multi-pulsar analysis of \S\ref{sec:Methods}. Note that the treatment of the white noise parameters is identical to that of \citet{Laal:2023etp}. No analytic conditional distribution can be found for the white noise parameters, so we must perform a short Metropolis-Hastings MCMC exploration to find a fair draw from the white noise conditional posterior during each loop of the Gibbs sampling. Although we could indeed sample white noise parameters in a global PTA analysis, as previously mentioned, we typically fix these to values determined from single-pulsar analyses. Hence in the following, we assume these are fixed by previous analyses, but the generalization of the Gibbs algorithm to sampling all parameters is, in theory, possible through the methodology implemented and tested previously in \citet{Laal:2023etp}.
\begin{itemize}[leftmargin=*]
    \item[]\textbf{Step 1}: Make an initial guess of $\Phi$, denoted by $\Phi_0$. This can be achieved practically by guessing an initial spectrum of PSD values for each pulsar to occupy the diagonal of $\Phi$. For the off-diagonal elements--- i.e., $\Phi_{IJ}$ where $I \neq J$--- the diagonal elements modulated by relevant values of the Hellings and Downs \citep{hd83} (HD) curve are an adequate initial guess.
    \item[]\textbf{Step 2}: An estimate of the vector $\bm{b}_0$ conditioned on $\Phi_0$ is drawn using \autoref{bgivenrho}.
    \item[]\textbf{Step 3}: An update to $\Phi$ is given by drawing a matrix $\Phi_1$ conditioned on $\bm{a}_0$ from $\bm{b}_0$ using \autoref{rhogivenb}.
\end{itemize}
\textbf{Steps 2} and \textbf{3} are then repeated in a loop until a desired level of convergence is reached for the Markov chains of all of the model parameters. The most common measures of convergence are \textit{effective sample size} (ESS) \citep{gelmanbook, arviz_2019} and the rank-normalized Gelman-Rubin $\widehat{\mathrm{R}}$ diagnostic \citep{gelmanrubin}. Both quantities measure the degree of statistical stationarity of the chain of samples that form a posterior probability distribution. More specifically, ESS measures the number of independent samples in a chain of samples (the higher the better) while $\widehat{\mathrm{R}}$ partitions the collective samples into multiple sub-chains and compares the variation within each sub-chain to variations across different sub-chains (the closer to 1 the better). We will use ESS to assess the convergence of the Markov chains that we generate in \S\ref{sec:results}.

In practice, since we are sampling large numbers of parameters, it is necessary to impose some constraints on certain values. Constraining $\Phi_{II}$ is necessary to avoid exploring parts of the parameter space that have no support under the prior. While \citet{constrainedwishart} introduce a method of constraining the diagonals of a Wishart distribution, the method is very computationally challenging and cannot be applied to Inverse-Wishart distributions easily as the proposed method involves a layer of iterative computation in which costly matrix products need to be performed. Instead, we use \emph{rejection sampling} (i.e., rejecting samples outside a certain fixed boundary) as a way of forcing the $\Phi_{II}$ parameters to be confined within $10^{-11}$ and $10^{-4}$ (units of $\text{seconds}^2$) for every frequency bin. With this practical constraint imposed during the Gibbs algorithm above, we now refer to the entire routine as \textit{Multi-pulsar Gibbs Sampling} (\texttt{MG}) from hereon.

The parameter dimensionality tackled by \texttt{MG} is daunting. A real PTA dataset with $100$ pulsars in which $30$ frequency-bins are used to model low-frequency red noise process has $6,000$ modes in $\bm{b}$, and $166,500$ covariance parameters in $\bm{\Phi}$. Without Gibbs sampling, the inclusion of all these parameters would be an even more daunting exercise, even with gradient-based MCMC methods like No-U-Turn Sampling \citep{Hoffman2011TheNS}. Current analyses by PTAs get around this obstacle by assuming a parameterized functional form for the elements of $\Phi$ and the inter-pulsar correlation structure, reducing the size of the parameter space significantly. However, our goal here is to produce posterior samples that are as agnostic as feasible 
 and allow such models to be imposed using \texttt{MG} samples in post-processing. We will discuss this point in \S\ref{sec:Resamp}.
\subsection{\label{Implementation Details}Implementation details}
\autoref{fig:flowchart} showcases a map of the computational tasks at each iteration of Gibbs sampling. To move from each node to another, the operation specified by the text surrounding the relevant arrow is performed on the quantity that the arrow originates from which results in the quantity that the arrow points towards. To estimate $\bm{a}$ from $\Phi^{-1}$, first the matrix $\Sigma^{-1}$ is constructed by following \autoref{sigma_Def}. Then, the upper-Cholesky-decomposition of $\Sigma^{-1}$, denoted by $U$, is calculated and used within two linear-algebra solvers to eventually draw a random set of $\bm{a}$ from the multivariate Gaussian distribution of \autoref{bgivenrho}. Note that $Z$ is a set of standard Gaussian random numbers required to sample from a multivariate normal distribution using the well-known \emph{affine transformation} technique\footnote{The affine transformation states that to sample from $N\left(\bm{\mu}, \Sigma \right)$, one can sample from $Z = N\left(0, \mathbb{I} \right)$ and find $\bm{\mu} + \sqrt{\Sigma}Z$, where $\sqrt{\Sigma}$ is a square-root factorization of $\Sigma$.}. To continue the flow, the scale-matrix $S$ is constructed for each frequency-bin using \autoref{scalemat}. For the sole purpose of numerical stabilization of $S$, first, the correlation version of $S$, denoted by $\hat{S}$, is calculated\footnote{The correlation version of any covariance matrix is found by element-wise division of the covariance matrix by the matrix constructed from the outer-product of the square-root of its diagonals.}. Second, for each pulsar and at each frequency-bin, a random number between $10^{-8}$ and $10^{-5}$ is drawn and added to the diagonals of the correlation matrix resulting in the matrix $\hat{S}^{+}$. The reason for performing such perturbations is entirely for numerical stability. Note that using numbers lower than $10^{-8}$ makes $\hat{S}^{+}$ to have a very high condition-number while using numbers larger than $10^{-5}$ changes the value of the scale-matrix along its diagonals by more than $10^{-5} \times 100=0.001$ percent. Furthermore, $\hat{S}^{+}$ is used to reconstruct $S$ which is now more numerically stable since it is cured of its low-rank deficiency. Finally, to complete the iteration and obtain $\Phi$ and $\Phi^{-1}$, a Bartlett decomposition is performed using the standard-Wishart random $n_p \times n_p$ matrix $A$ and the upper-Cholesky-decomposition, denoted by $P$, of the numerically stable matrix $S$. If the diagonals of the estimated $\Phi$ is not bounded by what the prior demands, the $\Phi$ matrix is discarded and another Bartlett decomposition using a completely different $A$ is performed.
\subsection{\label{challenges} Implementation challenges}
\begin{figure*}
    %\centering
\usetikzlibrary{shapes.geometric,arrows.meta}
\usetikzlibrary{positioning}
\usetikzlibrary{calc}
\tikzstyle{diam} = [diamond, aspect=2, draw, fill=red!40, text width=2em,text centered ]
\tikzstyle{diamg} = [circle, aspect=2, draw, fill=green!40, text width=2em,text centered ]
\tikzstyle{block} = [rectangle, draw, fill=blue!20, text width=8em,text centered, rounded corners, minimum height=2em ]

\tikzstyle{firstblock} = [rectangle, draw, fill=blue!20, text width=9.5em,text centered, rounded corners, minimum height=2em ]

\tikzstyle{trap} = [ellipse, draw, fill=gray!20, text width=10em,text centered, rounded corners, minimum height=2em ]
\tikzstyle{line} = [draw, -latex]

\tikzstyle{trapsmall} = [ellipse, draw, fill=gray!20, text width=5.4em,text centered, rounded corners, minimum height=2em ]
\tikzstyle{line} = [draw, -latex]

\centering
\begin{tikzpicture}[node distance=1.5cm and 2cm]

    % nodes
    \node [firstblock] (phi) {start: $\Phi^{-1}$} ;
    \node [diamg, right=of phi, xshift=2cm] (sigma) {$\Sigma^{-1}$} ;
    \node [diam, below=of sigma, yshift=1cm] (U) {$U$} ;
    \node [diamg, right=of U, xshift=1cm] (ahat) {$\hat{\bm{a}}$} ;
    \node [diamg, left=of U, xshift=-1.5cm] (mu) {$\bm{\mu}$} ;
    \node [block, below=of U, yshift=-0.2cm] (a) {$\bm{a} = \bm{\mu} + \hat{\bm{a}}$} ;
    \node [diamg, below=of a, yshift=1cm] (s) {$S$} ;
    \node [diamg, below=of s, yshift=1cm] (shat) {$\hat{S}$} ;
    \node [diamg, below=of shat, yshift=1cm] (shatplus) {$\hat{S}^{+}$} ;
    \node [diamg, below=of shatplus, yshift=1cm] (smod) {${S}^{+}$} ;
    \node [diam, left=of shatplus, xshift=-2.78cm] (P) {$\Upsilon$} ;
    \node [diam, above=of P, yshift=3cm] (X) {$\Psi$} ;
    \node [diam, left=of smod, xshift=-2.8cm] (Y) {$P$} ;
    %\node [trap, left=of P, xshift=2.5cm] (A) {$A$ (standard Wishart)} ;
    \node [trapsmall, left=of P] (A) {$A$ (standard\\ Wishart)} ;
    \node [trap, below=of U, yshift=1cm] (nu) {$Z$ (standard normal)} ;
    \node [block, above=of P, yshift=0.5cm] (diags) {$\Phi$, $\text{diag}\{ \Phi \}$} ;

    % lines
    \path [line] (phi) -- node[above] {add $T^TN^{-1}T$} node[below] {turn into $B$ matrix} (sigma) ;
    \path [line] (sigma) -- node[right] {take upper-Cholesky} (U) ;
    \path [line] (U) -- node[sloped, below] {solve $U\hat{\bm{a}}=Z$} (ahat) ;
    \path [line] (U) -- node[sloped, below] {solve $\Sigma^{-1} \bm{\mu}= T^TN^{-1}\bm{\delta t}$} (mu) ;
    \path [line] (ahat) |- node[right] {} (a) ;
    \path [line] (mu) |- node[right] {} (a) ;
    \path [line] (a) -- node[right] {perform outer-product of $\bm{a}$} (s) ;
    \path [line] (s) -- node[right] {find the correlation matrix of $S$} (shat) ;
    \path [line] (shat) -- node[right] {perturb diagonals of $\hat{S}$} (shatplus) ;
    \path [line] (shatplus) -- node[right] {reconstruct $S$ from $\hat{S}^{+}$} (smod) ;
    \path [line] (smod) -- node[fill=white] {take upper-Cholesky} (Y) ;
    %\path [line] (A.east) |- node[right] {} (X.west) ;
    \path [line] (Y) -- node[right] {solve $A \Upsilon= P$} (P) ;
    \path [line] (diags) -- node[fill=white] {if $\text{diag}\{ \Phi \}$ \textbf{is} bounded, solve $P \Psi= A$.} node [right] {} (X) ;
    \path[line] (X.north) -- node[left] {$\Phi^{-1} = \Psi \Psi^T$} (phi.south) ;
    \path [line] (nu) -- node[right] {} (U) ;
    \path [line] (A.south) |- node[right] {} (Y.south) ;
   \path [line] (P) -- node[fill=white] {$\Phi = \Upsilon ^T\Upsilon $} (diags) ;
    % \path [line] (diags.west) |- node[right, anchor=west,rotate=90,fill=white] {if $\text{diag}\{ \Phi \}$ not bounded} (Y.west) ;
   \path [line] (diags.south) edge[bend right=40] node[above, align=left, rotate=90, text width=2.5cm, text height=10em] {if $\text{diags}\{ \Phi \}$ \textbf{is not} bounded \vspace*{8pt} } (Y.west) ;
\end{tikzpicture}
    \caption{A schematic representation of the computational tasks carried at each iteration of our multi-pulsar Gibbs sampling. The quantity that the arrow is being originated from is used within the operation indicated by the text surrounding the arrow to result in the quantity that arrow is pointing towards. The color and shape coding of the blocks is to help distinguish the calculated quantities from each other. The blue rectangles contain the quantities that must be saved before the next iteration. The green circles contain auxiliary quantities that connect the different pieces together. The red diamonds represent the nodes where the numerical instability of various quantities might prevent the sampling iteration from completing successfully. And finally, the gray ellipses are the nodes where random draws from standard normal and Wishart distributions are made.}
    \label{fig:flowchart}
\end{figure*}
Beyond the practical constraints imposed on $\Phi_{II}$ discussed earlier, there are non-trivial challenges to the successful implementation of \texttt{MG}.
Except for $n_p \leq 2$ cases, a square-root factorization of the scale-matrix $S$ is not theoretically possible as $S$ is a rank-$2$ square matrix by construction. Therefore, in practice, in order to sample from an Inverse-Wishart distribution, $S$ must be cast into a positive-definite matrix that is factorizable as a product of a matrix and its transpose (e.g., a Cholesky decomposition). As stated in \S\ref{Implementation Details}, one can perturb the diagonals of the correlation matrix $\hat{S}$ by a very small amount to make $\hat{S}$, and hence $S$, full-rank and numerically stable. By stable, we are specifically referring to $(1)$ making $S$ to have a numerical Cholesky decomposition that is adequately accurate in reconstructing $S$ (i.e., $S \approx P P^T$ for a Cholesky factorization $P$), and $(2)$ ensuring that the condition number of the resulting decomposition of $S$ is as low as possible. There is a delicate balance between $(1)$ and $(2)$ which is the most difficult aspect of implementing \texttt{MG}. Other factorization techniques such as pivoted-Cholesky, truncated-SVD, and LDL decomposition \citep{pivchol} were tested with less optimal results, as their output is often a factorization of $S$ that is nearly a singular matrix.

To further illustrate these challenges, we briefly revisit the case of $n_p = 2$ outlined in \S\ref{sec:Two-pulsar}. The lower-Cholesky decomposition of $S=PP^T$ exists theoretically as
\begin{align}
P = \frac{1}{\sqrt{(a_1^c)^2 + (a_1^s)^2}} \left( \begin{matrix}
   (a_1^c)^2 + (a_1^s)^2 & 0  \\
    a_1^c a_2^c + a_1^s a_2^s & \left\| a_1^c a_2^s - a_1^s a_2^c \right\|  \\
\end{matrix} \right).
\end{align}
Nevertheless, this cannot be used practically within our Gibbs sampling routine as $P_{22}$ is often too close to being zero, especially for larger frequency-bins (there is one $P$ per frequency) where low-frequency signals are much smaller. Since $P$ is used in a linear-algebra solver routine to draw samples from \autoref{rhogivenb}, this affects the accuracy of results produced by the solver when $P_{22}$ gets too close to zero. Therefore, even for $n_p=2$, stabilization of $S$ is necessary and unavoidable.

\subsection{\label{Improving the convergence rate of Gibbs Sampling}Improving the convergence rate of Gibbs Sampling}
To help improve the stability and the convergence rate of \texttt{MG}, we introduce an \emph{empirical noise distribution jump} (ENDJ) after a certain number of Gibbs iterations. ENDJ is a common practice in PTA detection efforts when performing Bayesian inferences about models with unusually many ($> 100$) number of parameters. The idea behind ENDJ is simple: if one already has access to marginalized posterior probability distributions of a set of model parameters from another independent Bayesian inference routine, such distributions can be used as proposal distributions for a new Bayesian inference about the same set of model parameters. 

Even though samples from empirical posterior distributions such as from our ENDJ are model dependent and therefore not fully representative of the \texttt{MG} posterior distribution, they serve as effective proposal distributions. By accepting/rejecting with the standard Metropolis-Hastings ratio, we are assured to converge to the true MG posterior. The frequency at which the ENDJ draws are considered is set prior to the the start of the Gibbs sampling. In our implementation, the ENDJ draws are less frequent than the Gibbs sampling draws by at least two to four orders of magnitude.
\section{\label{sec:Resamp}Post-processing}

\texttt{MG} provides samples drawn from the joint posterior of $\bm{b}$ and $\Phi$ given a PTA dataset. The subsequent use of these samples in the inference of a parameterized model of $\Phi$ (e.g., a power-law spectrum) is non-trivial, as the posteriors on elements of $\Phi$ are covariant in extremely high dimensions. Nevertheless we provide some potential avenues through which this goal may be attained. These are not intended to necessarily be optimal strategies for post-processing the massive sample of posterior draws derived by an \texttt{MG} analysis, but rather a first exploration to motivate future work.

\subsection{Spectral re-fitting} \label{sec:spec_refit}

In \citet{fitting}, one-dimensional kernel density estimators (KDEs) are used to provide a representation of the MCMC-sampled marginal posterior distribution of the GWB's power spectrum at each frequency; this power spectrum can be obtained from a global PTA analysis or (by assuming a factorized form for the PTA likelihood) spectral estimation of each pulsar in turn. Although this approach neglects potential frequency covariance in the spectral estimates, the covariance was found to be small enough to minimally impact subsequent spectral fitting. In a similar way, \texttt{MG}'s estimate of the $\Phi$ matrix can be used to fit parametrized representations of the intrinsic red noise and common process contributions to the overall budget of low-frequency processes. 

Learning the posterior density of the $\Phi$ matrix from a finite set of posterior samples is an incredibly challenging task. The dimensionality of $\Phi$ is beyond the capabilities of 
multivariate KDEs, and may require machine learning techniques (e.g., deep normalizing flows). Hence, for the purposes of this paper, we only concern ourselves with learning the one-dimensional marginal posterior densities of $\Phi_{k;II}$ for each frequency and pulsar independently given the \texttt{MG} samples from the joint posterior.\footnote{This obviously neglects the inter-pulsar correlations, which we assume to be a subdominant source of information compared to the diagonal of $\Phi$. In the next subsection we provide a potential re-fitting technique for the inter-pulsar correlation entries.} Our re-fitting likelihood is then
\begin{equation}
    p( \bm{\delta t} | \bm{\theta} ) \propto \prod\limits_{k=1}^{n_f}{\prod\limits_{I=1}^{n_p} {p( \Phi_{k;II} = \Phi_{k;II}^M(\bm{\theta})  | \bm{\delta t} )}}  \label{KDELIK},
\end{equation}
where $p( \Phi_{k;II} | \bm{\delta t} )$ is the estimated one-dimensional marginal posterior of $\Phi$ at frequency $k$ in pulsar $I$. This posterior density is evaluated at $\Phi_{k;II}^M$, which is a model of $\Phi_{k;II}$ over all frequencies and pulsars, with parameters $\bm{\theta}$. For this model we adopt $\Phi_{k;II}^M = \kappa_{k;I} + \rho_k$, where $\kappa_{k;I}$ is the power-spectral-density of intrinsic red noise in each pulsar (which we model as independent power-law functions with two parameters per pulsar), and $\rho_k$ is the power-law power-spectral density of the common (correlated) process, with two global parameters. The parameter vector $\bm{\theta}$ is then of length $2n_p+2$, corresponding to the union of all intrinsic red-noise and common process power-law parameters. As in \citet{fitting}, we fit bandwidth-optimized KDEs to represent the marginal posterior densities of $\Phi$, allowing us to use \autoref{KDELIK} inside an MCMC inference procedure to obtain posterior samples on the parameters $\bm{\theta}$ of our model.

We note that even though we do not explicitly model the inter-pulsar correlation information in \autoref{KDELIK}, marginalization over all valid correlations is implicit in the posterior probability density of each of the $\Phi_{k;II}$ parameters. This is a subtle point: in \autoref{KDELIK} we are modeling the auto-spectral density of a general common correlated process, in which we marginalize over all inter-pulsar correlation signatures that are consistent with our data. This is thus different from the common uncorrelated red noise (CURN) model featured in \citet{15yr} and other papers. In the CURN model all inter-pulsar correlations have been assumed to be zero, whereas we have marginalized over all possible correlation signatures.

\subsection{Fitting cross-correlations} \label{sec:crosscorr_fit}

To re-fit the inter-pulsar covariance entries of $\Phi$ (i.e., $\Phi_{IJ}$ where $I \neq J$), an obvious approach is a $\chi^2$ minimization between the posterior draws of $\Phi_{IJ}$ and a parametrized model. By doing so for each $\Phi$ posterior draw, we derive a distribution of best-fit correlation parameters over the $\Phi$ posterior; this is akin to the approach taken in the noise-marginalized optimal statistic \citep{NMOS,OSperFreq}. We stress that this is the simplest approach that one could take, and is only an approximation as the one-dimensional marginal posteriors on $\Phi_{k;IJ}$ are not Gaussian distributed as shown in \S\ref{sec:Two-pulsar}. With these caveats, the $\chi^2$ statistic that we test as a re-fitting function is
\begin{equation}
    \chi^2_i = {\sum\limits_{k}^{n_f}} {\sum\limits_{I, J<I}} \left( \Phi^i_{k;IJ} - \left[ \frac{( f_k / f_\mathrm{yr} )^{3-\gamma}}{12\pi^2 f_k^3} \right] A^2\Gamma _{IJ}^{\mathrm{HD}} \right) \label{chifit}.
\end{equation}
where the index $i$ is over the set of $\Phi$ posterior samples drawn using \texttt{MG}. The term being subtracted from $\Phi_{IJ}$ is a power-law model of the GWB's cross-spectrum, in which $\gamma=13/3$ to represent the expected spectral shape given by a population of supermassive black-hole binaries, and $A$ is the GWB's characteristic strain amplitude at a frequency of $f_\mathrm{yr}=1/\mathrm{year}$. The template correlation pattern $\Gamma_{IJ}^{\text{HD}}$ is the HD curve, expected for a statistically isotropic distribution of GWB intensity. To distinguish estimates of $A_{\text{GWB}}$ obtained from minimizing \autoref{chifit} with respect to $A$ from those obtained through a Bayesian scheme using \autoref{KDELIK}, we use a ``hat'' on $A_{\text{GWB}}$ as in $\hat{A}_{\text{GWB}}$.

\begin{figure}
    %\centering
\includegraphics[width=\linewidth]{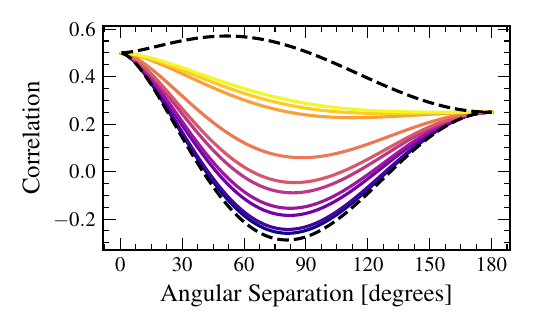}
    \caption{The generalized transverse (GT) parametrization of inter-pulsar correlations featured in \autoref{cross_corr_model}. The dashed lines denote the envelope of correlation curves given by $\tau\in U[-1.5,1.5]$, where solid colored curves correspond to particular random draws of $\tau$. The more positive the value of $\tau$, the closer the resulting pattern is to the top dashed bound. The HD curve is a particular case of this family of correlations corresponding to $\tau=-1$.}
    \label{fig:gtshape}
\end{figure}

We can also adopt a parameterized model of the correlation signature rather than merely assume that it is the HD curve. Moreover, we can do this at any given frequency bin $k$, to characterize the frequency-dependent correlation structure of the GW signal. A simple one-parameter model to fit is the so-called generalized transverse (GT) overlap reduction function \citep{enterprise-ex, nimaythesis} given by
\begin{equation}
    \Gamma_{k;IJ}^\mathrm{GT} =  \frac{3}{4}\left( 1-\tau_k  \right)x_{IJ}\ln(x_{IJ}) - \frac{x_{IJ}}{4} +\frac{1}{2}, \label{cross_corr_model}
\end{equation}
such that $\Phi_{k;IJ} = \Gamma_{k;IJ}^\mathrm{GT} \rho_{k} $, where $I\neq J$, $x_{IJ} = (1-\cos\xi_{IJ})/2$, and $\xi_{IJ}$ is the angular separation between pulsars $I$ and $J$. The parameters describing the power-spectral-density and shape of the correlations at each frequency are $\rho_{k}$ and  $\tau_k$, respectively. The GT model includes the HD curve as a special case where $\tau=-1$. \autoref{fig:gtshape} shows the shape of the GT function for different values of $\tau$ drawn uniformly between $-1.5$ and $1.5$. 

To perform the $\chi^2$ minimization itself, a simple curve-fitting optimizer is sufficient to find the distribution of best-fit $A^2$ in \autoref{chifit}. We find the best performance with the nonlinear least-squares Levenberg-Marquardt algorithm (more commonly known as the LM method) \citep{lm} to determine the best-fit $\rho_k$ and $\tau_k$ in \autoref{cross_corr_model}. 

\section{\label{sec:sims} Simulations \& Results}

We assess the performance of \texttt{MG} and the post-processing schemes outlined earlier on several simulated PTA datasets. Our aim is not to perform a comprehensive study of \texttt{MG} or to gauge its performance on all possible types of simulated and real data sets, but to introduce this new technique and show its basic capabilities. The current standard detection techniques were built over many years and improved in both efficiency and quality over time. We expect the same progression for \texttt{MG}. In the following section, we apply \texttt{MG} on two sets of simulated data and report on the results of post-processing compared to standard techniques. We defer a more comprehensive study of the output of \texttt{MG} to future work.
 
\subsection{\label{sec:ccrn} Fiducial model}
\begin{figure*}%[!ht]
\centering
\subfloat{\includegraphics[width=0.49\textwidth]{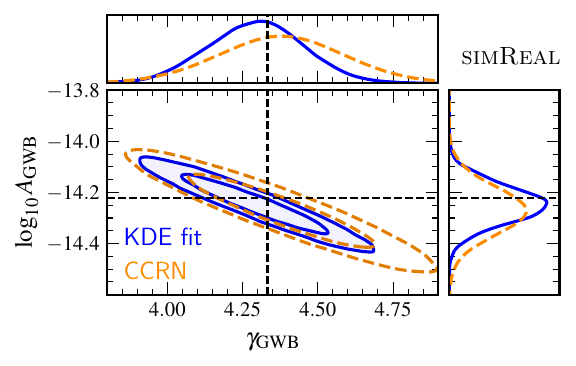}}
\subfloat{\includegraphics[width=0.49\textwidth]{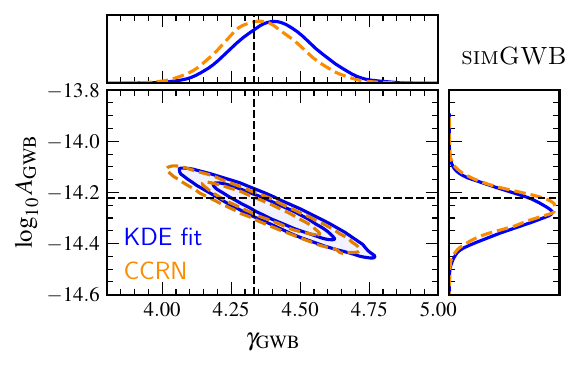}}\\
\subfloat{\includegraphics[width=0.49\textwidth]{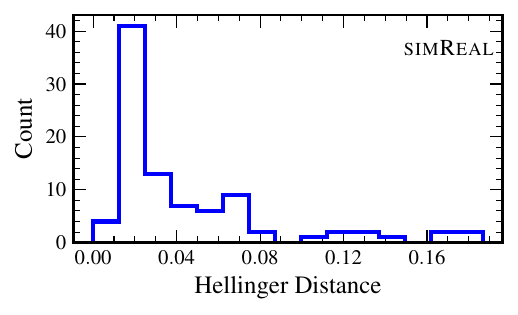}}
\subfloat{\includegraphics[width=0.49\textwidth]{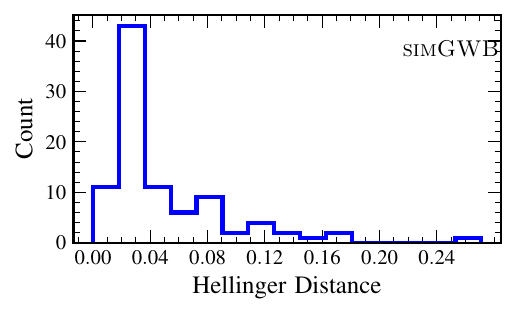}}
\caption{Plots depicting the posterior probability density for the GWB amplitude and spectral-index obtained using CCRN (dashed orange) and through KDE fit to $\Phi_{II}$ of MG (solid blue) for \textsc{simReal} in the top-left panel and for simGWB in the top-right panel. In the bottom panels, the degree of consistency between the two technique's posterior distributions is shown. The histogram ranges over all common model parameters: $\left\{ \theta _{1},\cdots ,\theta _{{{n}_{p}}}, {{\gamma }_{\text{GWB}}},{{\log }_{10}}{{A}_{\text{GWB}}} \right\}$. Overall, despite the fundamental differences between the two techniques, the distributions are consistent.}
\label{fig:kdefits}
\end{figure*}
The most fair comparison of \texttt{MG} to standard detection pipelines would be if the latter were able to infer the cross-power-spectral-density between each pulsar pair. However, as mentioned earlier, the dimensionality of this problem is too high for standard techniques based on Metropolis-Hastings MCMC sampling to confront. Therefore, as a proxy for this in our standard detection pipeline (built with the PTA data analysis suite \texttt{enterprise} \citep{enterprise, enterprise-ex}), we assume a flexible model for the inter-pulsar correlations in which these correlations are the same at all GW frequencies, but are allowed to vary in seven independent angular-separation bins, similar to Figure 1-(d) of \citet{15yr}.\footnote{The 7 bin locations are chosen to capture the maximum and minimum angular separations, the two zero crossings of the HD curve as well as the position of minimum correlation.} We call this a common cross-correlated red noise (CCRN) process; it captures some of the flexibility of \texttt{MG} while still running in a reasonably short amount of computational time. Spectra in this model are treated as power-law across frequencies for the CCRN process and for intrinsic pulsar red noise, with associated power-law amplitude and spectral index parameters. This model is therefore that of \autoref{eq:plaw_irn_hd}, with $\Gamma_{IJ} = \Gamma_{IJ}(\bm{c})$ modeled as a piecewise constant function with inferred values in different angular separation ranges.

The model has $2n_p+9$ parameters: an amplitude and spectral index describing power-law intrinsic red noise in each pulsar ($\gamma_I, \log_{10}A_I$), an amplitude and spectral index describing the power-law CCRN ($\gamma_\mathrm{CCRN}, \log_{10}A_\mathrm{CCRN}$), and seven normalized cross-correlation parameters $\bm{c}$, with each assigned to every pulsar pair whose angular separation lies in a given angular-separation bin. We place uninformative uniform priors on all $\bm{\theta}^{\text{CCRN}}$ parameters: 
\begin{align}
\begin{split}
   p\left( {{\gamma }_{I}} \right)&=\text{Uniform}\left( 0,7 \right), \\ 
  p\left( {{\gamma }_{\text{CCRN}}} \right)&=\text{Uniform}\left( 0,7 \right), \\ 
  p\left( {{\log }_{10}}{{A}_{I}} \right)&=\text{Uniform}\left( -18,-11 \right), \\ 
 p\left( {{\log }_{10}}{{A}_{\text{CCRN}}} \right)&=\text{Uniform}\left( -18,-11 \right),\\
 p\left(c \right) &=\text{Uniform}\left( -1,1 \right).
 \end{split}
\end{align}

Even though we consider the CCRN model as our fiducial model, an alternative is the aforementioned GT model from \autoref{cross_corr_model}. In a Bayesian GT analysis, we can vary and constrain $\tau_k$ simultaneously across all frequencies---since there is only one shape parameter per frequency---and still obtain converged MCMC chains within a reasonably short amount of computational time. This model has $2n_p+2+n_f$ parameters, with the same power-law spectrum priors as the CCRN model, and with priors on the GT parameters corresponding to $p(\tau_k)=\mathrm{Uniform}[-1.5,1.5]$, for all $k$. In cross-spectrum analyses, we use the result of GT modeling as our point of reference for what the standard techniques are capable of revealing about the shape of the correlations at a given frequency-bin.

In summary, we post-process the output of \texttt{MG} using the approaches outlined in \S\ref{sec:Resamp} in order to compare the performance of spectral characterization and cross-correlation reconstruction to the standard \texttt{enterprise} detection pipeline with the CCRN (and GT) model.

\subsection{\label{sec:sim data} Simulated data sets}

We create and analyze two different sets of simulated PTA data, each with $45$ pulsars. For this proof-of-principle study, we inject signals into, and recover signals from, the first five sampling frequencies of the datasets (i.e., $f_k = k/T_\mathrm{obs}$ with $k\in[1,5]$ and $T_\mathrm{obs}=20$~years). The two datasets are:
\begin{itemize}[leftmargin=*]
    \item[]\textsc{simReal}--- a realistic PTA dataset, based on the simulation pipeline from \citet{Astroforcast} in which the pulsar observation schedule, observing baseline, white noise level, and the intrinsic red noise spectral parameters are drawn from the NANOGrav 12.5-year dataset \citep{12p5timing}. The dataset has an overall timing baseline of $20$~years, but individual pulsars vary in length from $\sim10$ to $20$~years. A power-law spectrum GWB signal is injected with $\gamma_\mathrm{GWB}=13/3$ and $A_\mathrm{GWB}=6\times 10^{-15}$.

    \item[]\textsc{simGWB}--- a more theoretical PTA dataset, identical to simReal except for its white noise and observing baseline. For this data set, all pulsars have the same white noise RMS of $10$~ns, and all are observed for $20$~years. As in \textsc{simReal}, a power-law spectrum GWB signal is injected with $\gamma_\mathrm{GWB}=13/3$ and $A_\mathrm{GWB}=6\times 10^{-15}$.
\end{itemize}

\subsection{\label{sec:results} Results}

\subsubsection{\label{sec:q1} Characterizing the spectrum}

We use the techniques outlined in \S\ref{sec:spec_refit} to fit for the powerlaw model parameters of each pulsars' common and intrinsic red noise using the sampled output of \texttt{MG} for each of our two simulated datasets. We also analyze each simulated dataset directly with the CCRN model implemented in the standard \texttt{enterprise} detection pipeline, with the result being joint posterior distributions on all parameters of the model, $\bm{\theta}^\mathrm{CCRN}$, for each dataset obtained via both \texttt{MG} refitting and direct analysis.   

\autoref{fig:kdefits} shows the consistency between the refitting results and CCRN analyses. As is evident, both \texttt{MG}-refitting (solid blue) and CCRN (dashed orange) deliver comparable joint posteriors on $\{\log_{10}A_\mathrm{GWB},\gamma_\mathrm{GWB}\}$. More explicitly, for the \textsc{simReal} data set, \texttt{MG}-refitting gives ${{\log }_{10}}{{A}_{\text{GWB}}}=-14.25_{+0.07}^{-0.08}$ compared to an injected value of ${{\log }_{10}}{{A}_{\text{GWB}}}=-14.22$, and ${{\gamma }_{\text{GWB}}}=4.30_{+0.16}^{-0.15}$ compared to the injected value ${{\gamma }_{\text{GWB}}}=4.33$. Note that the reported values are median and $1$-$\sigma$ equivalent percentiles. Additionally, for the \textsc{simGWB} data set, \texttt{MG}-refitting gives ${{\log }_{10}}{{A}_{\text{GWB}}}=-14.27_{+0.08}^{-0.07}$ and ${{\gamma }_{\text{GWB}}}=4.41_{+0.14}^{-0.14}$.

To quantify any differences between the \texttt{MG}-refit and direct-CCRN analyses, we calculate \emph{Hellinger distances} \citep{Hell} between the resulting posterior distributions. The Hellinger distance measures the similarity between two probability distributions, bounded in the range $[0,1]$; the lower this value, the more consistent are the two distributions. More precisely, for two discrete probability distributions $P = (p_1, p_2, \ldots , p_r)$ and $Q = (q_1, q_2, \ldots , q_r)$, the Hellinger distance $H$ between $P$ and $Q$ is defined as 
\begin{equation}
    H = \frac{1}{\sqrt{2}} \sqrt{\sum\limits_{i=1}^{r}{{{\left( \sqrt{{{p}_{i}}}-\sqrt{{{q}_{i}}} \right)}^{2}}}}.
\end{equation} 
An intuition for Hellinger distances was given by \citet{fitting}, where two equal-variance Gaussian distributions whose means are offset by $1(2)$-$\sigma$ have Hellinger distances of $0.34(0.63)$.

The bottom panels of \autoref{fig:kdefits} show the distribution of Hellinger distances between \texttt{MG}-refit and direct-CCRN results computed for the one-dimensional marginal posteriors of each parameter in the CCRN model. Both simulated datasets exhibit small Hellinger distances that are well within an effective $1$-$\sigma$ discrepancy. More explicitly, for $\textsc{simGWB}$ and $\textsc{simReal}$ the median and the $1$-$\sigma$-equivalent percentiles for the Hellinger distance estimates are $0.03_{+0.07}^{-0.02}$ and $0.03_{+0.08}^{-0.02}$ respectively. Given the fundamental differences between the two techniques, specifically in how the spectrum fitting is performed for \texttt{MG} and the approximations made within, the consistency of the results is excellent. 
\subsubsection{\label{sec:q2}Characterizing the cross-spectrum} 

\begin{figure}
    %\centering
\includegraphics[width=\linewidth]{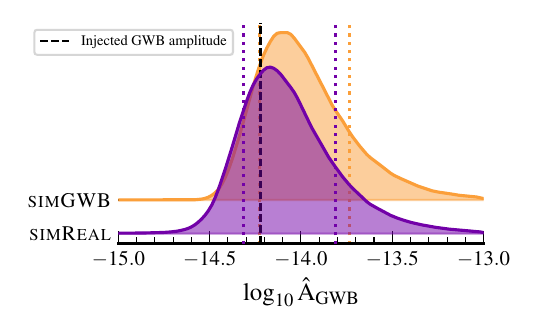}
    \caption{The distribution of $\log_{10}$ of the optimal estimator of $A_{\text{GWB}}$ found by applying the $\chi^2$ statistic of \autoref{chifit} on unique random draws from the posterior of $\Phi_{IJ}$. Note the use of ``hat'' on $A_{\text{GWB}}$ to emphasize the fact that the distributions are made from best-fit values. The lower distribution (purple) is from analyzing \textsc{simReal}, whereas the upper distribution (light orange) is from analyzing  \textsc{simGWB}. For both datasets, the injected GWB amplitude (black vertical dashed line) is $A_\mathrm{GWB} = 6 \times 10^{-15}$. The 1-$\sigma$ level of each distribution is identified with vertical dotted lines matching the color of their corresponding distribution. }
    \label{fig:chisqcross}
\end{figure}

\begin{figure*}[!ht]
\centering
\subfloat{\includegraphics[width =0.49\linewidth]{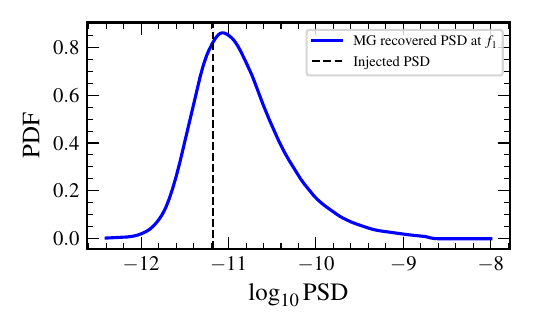}}
\subfloat{\includegraphics[width =0.49\linewidth]{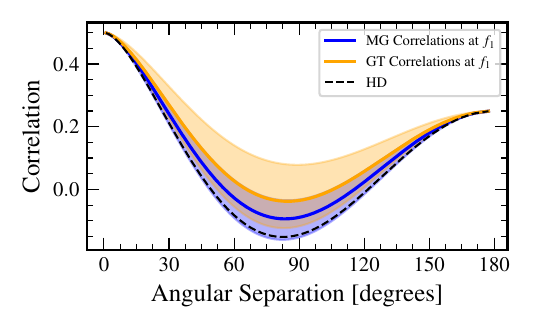}}
\caption{Left: a density plot of power-spectral density fits to the cross-correlation estimates of \texttt{MG} ($\Phi_{IJ}$ for $I \neq J$) at the first frequency-bin is depicted. Right: a plot of the reconstructed shape of the correlations (median and 1-$\sigma$ level) using fits to $\Phi_{IJ}$ is shown. Both plots are obtained through LM fits to random draws from \textsc{simReal}'s $\Phi_{1;IJ}$ posterior probability distributions. The fitted model follows \autoref{cross_corr_model} with only two parameters: $\log_{10}{\rho_1}$ for the power-spectral density and $\tau$ for the shape of the correlations.}
\label{fig:crosscorr}
\end{figure*}

We now use the techniques described in \S\ref{sec:crosscorr_fit} for fitting inter-pulsar--correlated models to the sampled output of \texttt{MG}. 
\autoref{fig:chisqcross} shows the result of applying the $\chi^2$ statistic of \autoref{chifit} to the joint posterior on all $\Phi_{IJ}$ elements (with $I\neq J$) that were sampled by \texttt{MG} on both sets of simulated data. To make \autoref{fig:chisqcross}, we randomly draw $2 \times 10^{5}$ unique samples from the $\Phi_{IJ}$ \texttt{MG} posterior, noting that these draws respect the measurement covariance between different pulsar pairs. For each draw, the value of $A_{\text{GWB}}$ that minimizes the $\chi^2$ statistic of \autoref{chifit} is estimated (denoted by $\hat{A}_{\text{GWB}}$); hence \autoref{fig:chisqcross} is a distribution of derived best-fit GWB amplitudes over the \texttt{MG} posterior. We can see that, in these cases, the distributions are consistent with the injected amplitude value to within the 1-$\sigma$ interval of the distributions.

Instead of assuming that the pattern of inter-pulsar correlations conforms to the HD curve, we now use the GT model described by \autoref{cross_corr_model}. The general procedure is outlined in \S\ref{sec:crosscorr_fit}; in this specific test we use the LM algorithm to find the best-fit power-spectral-density value, $\rho_1$ and GT shape parameter, $\tau$, at frequency $f_1 = 1/T_{\text{obs}}$. \autoref{fig:crosscorr} shows the result of fitting the GT model to $2 \times 10^{5}$ unique random draws from the $\Phi_{1;IJ}$ posterior of \textsc{simReal}. The LM algorithm is applied to these draws to find the best-fit model parameters. The left panel of \autoref{cross_corr_model} shows the uncertainty distribution on $\log_{10}{\rho_1}$, which is consistent with the injected value shown as a dashed vertical line. The best-fit $\tau_1$ estimates are used to create a posterior envelope of GT curves, shown as a solid median curve and shaded $1$-$\sigma$ region. The true correlation pattern---i.e., the HD curve---is shown as a dashed line that falls in the shaded envelope of the  reconstructed inter-pulsar correlation curves.

\section{\label{sec:IWdraws}No sampling--- approximating the GWB characterization using Inverse-Wishart distribution}

\begin{figure}
    %\centering
\includegraphics[width=\linewidth]{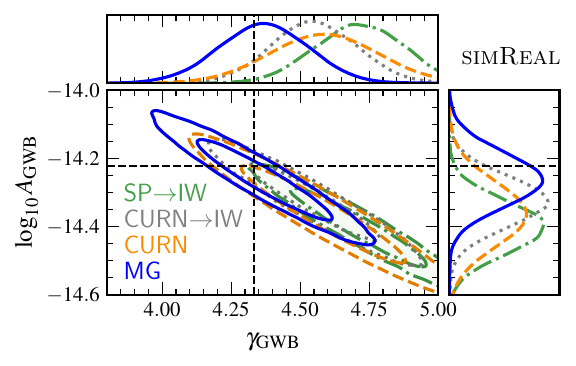}
    \caption{Density plots depicting the posterior probability density for the GWB amplitude and spectral-index obtained using $\texttt{MG}$ (solid blue), $\texttt{CURN} \to \texttt{IW}$ (dotted gray), $\texttt{SP} \to \texttt{IW}$ (dash dotted green) and standard CURN (dashed orange). The underlying data set is $\textsc{simReal}$. All but the standard CURN posterior probabilities are obtained via the fitting method described in \S\ref{sec:spec_refit} using the appropriate $\Phi_{II}$ generated samples. The only pair of consistent distributions are CURN's and $\texttt{CURN} \to \texttt{IW}$'s distributions evident by Hellinger distance of $0.12$ and $0.18$ for $\log_{10}A_{\text{GWB}}$ and $\gamma_{\text{GWB}}$ parameters respectively. }
    \label{fig:nosampling}
\end{figure}

\begin{figure*}
    %\centering
\includegraphics[width=\linewidth]{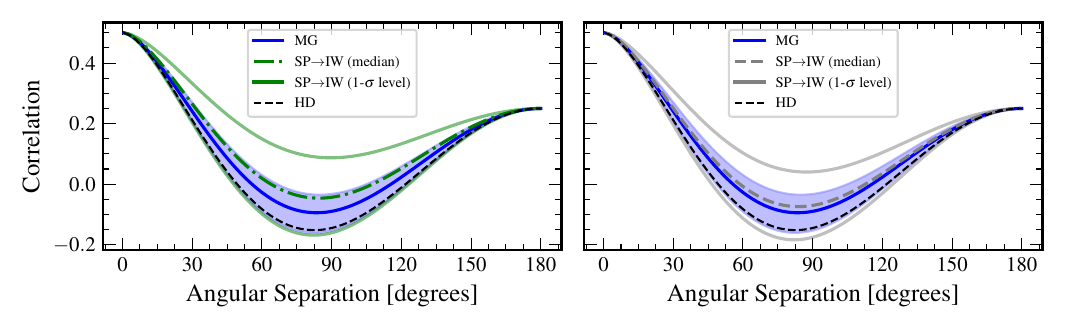}
    \caption{The reconstructed shape of the correlations by performing an LM fit according to \S\ref{sec:crosscorr_fit} using samples from $\texttt{MG}$ (blue), $\texttt{CURN} \to \texttt{IW}$ (gray), and $\texttt{SP} \to \texttt{IW}$ (green). Although varying in their median (indicated by inner curves) and 1-$\sigma$ level uncertainties (indicated by solid outer lines forming an envelope or shaded regions), all distributions overlap with the HD curve and each other signifying a level of consistency between the results.}
    \label{fig:nosampling_simReal}
\end{figure*}

\begin{figure}
    %\centering
\includegraphics[width=\linewidth]{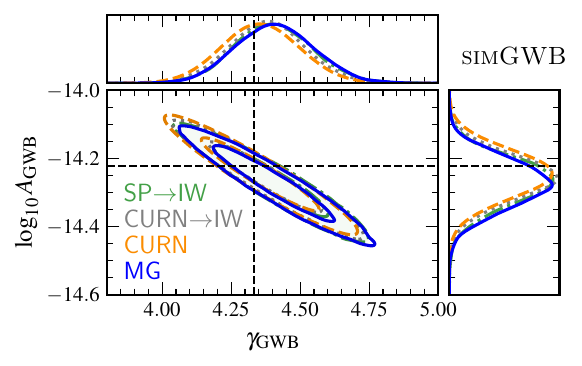}
    \caption{Density plots depicting the posterior probability density for the GWB amplitude and spectral-index obtained using $\texttt{MG}$ (solid blue), $\texttt{CURN} \to \texttt{IW}$ (dotted gray), $\texttt{SP} \to \texttt{IW}$ (dash dotted green) and standard CURN (dashed orange). The underlying data set is $\textsc{simGWB}$. All but the standard CURN posterior probabilities are obtained via the fitting method described in \S\ref{sec:spec_refit} using the appropriate $\Phi_{II}$ generated samples.}
    \label{fig:theothreepsds}
\end{figure}

The implementation challenges discussed in \S\ref{challenges} motivate efforts to approximate the output of \texttt{MG} sampling. This is a difficult task, as not only does the probability distribution of every single one of the many parameters need to be approximated correctly on an individual basis, but their covariance with respect to each other must also be correctly approximated. 

The approximation method we propose hinges on the idea that the posterior probability distribution of the Fourier coefficients that one obtains using \texttt{MG} could be similar to the ones obtained via existing standard methods of performing Bayesian inferences, i.e., the distribution has some overlap with one obtained using less--computationally-intensive standard techniques. Two standard methods that may yield such similar posterior probability distributions are $(1)$ single-pulsar analyses performed via Gibbs sampling \citep{Laal:2023etp}, and $(2)$ searches for a CURN process (as featured in \citet{15yr}) performed via a quasi-Gibbs sampling algorithm introduced in \citet{Gibbs0}. Note that a CURN model is identical to a CCRN model except for the pattern of inter-pulsar correlations; in the former, these are assumed to be zero at all frequencies. 

Since CURN modeling is done at the PTA (i.e., global) level, posterior samples of Fourier coefficients in a given pulsar and at a given GW frequency will reflect any measurement covariance with other pulsars and frequencies. Consequently, CURN modeling possesses a clear advantage over independent single-pulsar analyses in its ability to approximate an appropriate combination of Fourier coefficients, $\{a_{k;I}\}$, rather than merely estimate them from each pulsar in isolation. CURN is thus a great candidate model to use for estimates of Fourier coefficients as approximants of draws from \texttt{MG}.

We formalize our approximation method as follows. To use the result of single-pulsar analyses, we simultaneously draw $n_\text{draw}$ independent and unique samples from the posterior probability distribution of all pulsars' Fourier coefficients such that $n_\text{draw} \gg n_p$, which we further use to form
\begin{align}
{{S}_{k;IJ}}=\frac{1}{n_\text{draw}} \left[\sum\limits_{i=1}^{n_\text{draw}}{{{\left( a_{k;I}^{c} \right)}^{i}}{{\left( a_{k;J}^{c} \right)}^{i}}} + {{{\left( a_{k;I}^{s} \right)}^{i}}{{\left( a_{k;J}^{s} \right)}^{i}}}\right] \label{sample-avg}.
\end{align}
The condition $n_\text{draw} \gg n_p$ ensures that $(1)$ the resulting-scale matrix is positive-definite and $(2)$ no single-sample lies on or near the hyperplane spanned by the rest of the samples. 
This in turn guarantees a successful application of the Bartlett decomposition technique to analytically draw $\Phi_k$ samples conditioned on $S_k$ using \autoref{rhogivenb}. Given $n_p=45$ for the simulated data sets we considered in this work, $n_\text{draw} = 1000$ is adequate. We repeat this process (forming a random $S_k$ and drawing from \autoref{rhogivenb}) until the resulting distributions for the elements of $\Phi$ matrix is stationary\footnote{The more iterations done to form a distribution for $\Phi$, the less important the exact value of $n_\text{draw}$ becomes. One can set $n_\text{draw} = n_p$ and obtain the same distribution given sufficient repetition of the procedure.}. 

To preserve the correlations across frequency-bins and pulsars between the Fourier coefficients uncovered by CURN, we take a different approach. We simply draw randomly (one single draw) from the posterior probability distribution of the CURN's Fourier coefficients and form a $S_k$ matrix based on \autoref{scalemat}. We then cure $S_k$ of its low-rank deficiencies using the same methodology as used within $\texttt{MG}$. The resulting fully-ranked $S_k$ is hence ready to be used within a Bartlett-decomposition to yield estimates of the $\Phi$ matrix. Similar to the single-pulsar case, we repeat this process until the distributions of the elements of $\Phi$ matrix are all stationary. For ease of reference, we refer to the approximation method that uses CURN's Fourier coefficients as $\texttt{CURN} \to \texttt{IW}$ and the one that uses single-pulsar Fourier coefficients as $\texttt{SP} \to \texttt{IW}$\footnote{$\texttt{SP}$ and $\texttt{IW}$ are shorts for single-pulsar and Inverse-Wishart respectively.}. 

\autoref{fig:nosampling} and \autoref{fig:nosampling_simReal} showcase the effectiveness of the presented approximation method for \textsc{simReal} data set. For \autoref{fig:nosampling}, we use the techniques outlined in \S\ref{sec:spec_refit} to fit for the powerlaw model parameters of each pulsars' common and intrinsic red noise using the $\Phi_{II}$ samples generated from \texttt{MG}, $\texttt{CURN} \to \texttt{IW}$, and $\texttt{SP} \to \texttt{IW}$. As evident by the figure, standard CURN analysis and $\texttt{CURN} \to \texttt{IW}$ are consistent with each other (Hellinger distances of $0.12$ and $0.18$ for $\log_{10}A_{\text{GWB}}$ and $\gamma_{\text{GWB}}$ respectively) while the other two techniques yield different distributions. We attribute such discrepancies to the model-dependency of the Fourier coefficient recovery to the priors used on the red noise PSD parameters as well as assumptions used within each noise model. Such model-dependency is especially noticeable in analyses of realistic data sets as the prior dominates the likelihood at certain regions of the parameter space. Lastly, \autoref{fig:nosampling_simReal} illustrates how the LM fitted first frequency-bin correlations obtained from the approximation methods and $\texttt{MG}$ fair against each other. Remarkably, the three methods are consistent with each other in describing the correlations with the $\texttt{SP} \to \texttt{IW}$ method having the widest 1-$\sigma$ level uncertainty region as well as the most deviant median away from the injected HD correlations. $\texttt{MG}$'s correlations are the most constrained and closest to the HD curve.

Note that the success of the presented approximation method is dependent on the closeness of \texttt{MG}'s Fourier coefficients to the ones obtained from CURN and single-pulsar analyses as mentioned previously. For theoretical simulated data sets with strong red noise signal and ideal timing conditions, this is not an issue as Fourier coefficients' recovery is not much sensitive to the Bayesian models used to infer them. Hence, single-pulsar analyses would yield the same distribution of Fourier coefficients as \texttt{MG} or CURN would. However, for reasons yet to be fully understood, this becomes an issue with realistic data sets: single-pulsar results disagree with PTA-level analyses significantly in both PSD and Fourier coefficient recovery of the red noise of many pulsars.

\autoref{fig:theothreepsds}, made for $\textsc{simGWB}$, proves this claim. Similar to \autoref{fig:nosampling}, this figure shows a comparison between the recovered powerlaw model parameters of each pulsars' common red noise using fitted samples from \texttt{MG}, $\texttt{CURN} \to \texttt{IW}$, $\texttt{SP} \to \texttt{IW}$ and samples directly from a standard CURN analysis. Since $\textsc{simGWB}$ is free of the complications that make $\textsc{simReal}$ a realistic data set (i.e., pulsars having very different observational baselines and possessing significant white noise levels), the Fourier coefficients recovered using single-pulsar, CURN, and $\texttt{MG}$ are all very similar to each other resulting in different analyses to yield very similar distributions for $\log_{10}A_{\text{GWB}}$ and $\gamma_{\text{GWB}}$. Lastly, a direct comparison between the left panel of \autoref{fig:nosampling} and \autoref{fig:theothreepsds} further highlights one point: if one has access to what one believes is the best estimates of Fourier coefficients, then performing the Gibbs sampling is unnecessary and the approximation methods discussed in this section are sufficient in characterizing the GWB.  

\section{\label{sec:conclusion}Conclusion and Future Work}

In this work, we outlined the most agnostic, per frequency, Bayesian search for a low-frequency red noise process in a PTA data set using a Jeffrey's-like multivariate conjugate prior for the red noise covariance matrix. Every single independent entry in the red noise covariance matrix at a specific frequency-bin is treated as a model parameter for which a marginalized posterior probability distribution can be found via Gibbs sampling. We refer to the sampling and the noise modeling presented in this work as $\texttt{MG}$, short for multi-pulsar Gibbs sampling routine. Despite being fast and scalable, $\texttt{MG}$'s novelty lies in its generality and unique approach to the PTA gravitational wave data analysis problem rather than its computational efficiency. Knowing the extent of the dependency of the GWB characterization to the details of the Bayesian models describing each pulsar's red noise \citep{Hazboun_2020}, $\texttt{MG}$'s agnostic noise modeling enables for a unique way to probe a GWB signal in a PTA data set.

As shown in this work, spectrum fitting techniques using one-dimensional kernel-density-estimators could be used to refit the red noise auto-spectrum samples generated by $\texttt{MG}$ and obtain posterior probability distributions for both common correlated and intrinsic uncorrelated red noise model parameters for each pulsar. Furthermore, we show examples of how one can perform a cross-correlation-only analysis using the cross-spectrum samples generated by $\texttt{MG}$. The first example involves using a $\chi^2$ statistic to find a distribution for the optimal estimator of the amplitude of a powerlaw model with fixed spectral index and fixed correlations at HD values. The powerlaw model whose amplitude is estimated using the $\chi^2$ statistic describes the functional form of the cross-correlation power-spectral-density of the GWB, akin to the \emph{noise marginalized optimal statistic} technique \citep{OSperFreq, NMOS}. Additionally, we show how to perform an LM fit to fit a specific model to the cross-correlation samples of $\texttt{MG}$; hence, perform inference about the model parameters that characterize the correlations induced by a GWB signal. 

One notable feature of the noise modeling of $\texttt{MG}$ is that sampling is only required if the Fourier coefficients and the red noise covariance matrix are both simultaneously unknown. Since Fourier coefficients are pulsar-level parameters, besides $\texttt{MG}$, there are other methods to perform Bayesian inference about them and obtain posterior probability distributions for such coefficients that approximate what $\texttt{MG}$ would have recovered. In this paper, we show how using the Fourier coefficients recovered via single-pulsar and CURN (search for common uncorrelated red noise) analyses can be used as approximants of $\texttt{MG}$'s Fourier coefficients, in effect, reducing the problem of Bayesian inference of the red noise covariance matrix to analytic draws from a specific Inverse-Wishart distribution. This approximation method allows for single-pulsar analyses to be used to characterise both auto and cross-spectrum of a GWB: a feat most useful for \emph{advanced noise modeling} efforts \citep{15yrANM}. Since the inclusion of complicated customized noise models of individual pulsars in a global PTA-level analysis is computationally difficult using standard techniques, one can adopt the methodology presented in this paper to observe the effects of the advanced noise models on GWB's characterization without the need to perform a computationally expensive global PTA-level analysis.

The generality of the $\texttt{MG}$'s noise modeling comes at a price. The price is numerical instability of linear algebra operations that need to be carried to complete every iteration of Gibbs sampling. In this paper, we show that the application of our current implementation of $\texttt{MG}$ on both realistic and theoretical data sets is consistent with the injected signal as well as the results of the standard analysis aimed at characterizing the auto and cross-spectrum of the GWB. Nevertheless, to prepare $\texttt{MG}$ for analyzing real data set, we will need to improve on our current implementation of $\texttt{MG}$ ensuring its numerical stability and scalability at all times during the sampling steps. The improvements are needed as reaching adequate levels of convergence for the Fourier coefficients becomes more and more difficult with increasing number of frequency-bins and pulsars with mostly white noise dominated data. The scalability of $\texttt{MG}$ to include more pulsars and frequency-bins may be resolvable by using modern GPU-based array-oriented numerical computation software such as $\texttt{JAX}$ \citep{jax2018github} to reduce the computational cost of steps shown in \autoref{fig:flowchart}. Additionally, we speculate that the most effective solution to resolve the numerical problems of $\texttt{MG}$ is possible through constrained draws from the multivariate normal distribution of \autoref{bgivenrho} to prevent the Gibbs sampler from getting stuck at regions of parameter space that correspond to unreasonably weak ($<10^{-8}$ seconds) Fourier coefficient estimate.  

Besides improving on the implementation of $\texttt{MG}$, multivariate density estimation of the entire red noise covariance matrix needs to be developed to be able to fit an arbitrary parameterized model of a red noise covariance matrix to the output of $\texttt{MG}$. Currently, the fitting can be performed on the auto or cross-spectrum of the red noise covariance matrix separately, but not on both simultaneously. Given the enormous dimensionality of the red noise covariance matrix, machine-learning techniques need to be employed to achieve such goal.

\subsection{\label{sec:software}Software}
The multi-pulsar Gibbs sampling code \citep{Laal_PANDORA_2025} takes advantage of the functionalities provided by  \texttt{Numpy} \citep{numpy},  \texttt{Scipy}, \citep{scipy}, \texttt{Ceffyl} \citep{fitting}, \texttt{enterprise} \citep{enterprise} and  \texttt{enterprise-extensions} \citep{enterprise-ex}. Python packages  \texttt{matplotlib} \citep{plt}, and \texttt{chainconsumer} \citep{chainconsumer} have been used for generating the scientific plots in this paper. The simulated data sets can be found in \citet{Laal_PANDORA_2025}.
\begin{acknowledgments}
We thank the anonymous referee for their helpful feedback which improved the quality of this work. We thank Bernard Whiting for pointing out plotting errors in some of the figures featured in this work. We thank our colleagues in NANOGrav for fruitful discussions and feedback during the development of this technique. The work of N.L., W.G.L, X.S., and S.R.T was supported by the NANOGrav NSF Physics Frontier Center awards \#2020265 and \#1430284. X.S acknowledges the support from the George and Hannah Bolinger Memorial Fund at Oregon State University. S.R.T acknowledges support from NSF AST-2007993, and an NSF CAREER \#2146016. N.L is supported by the Vanderbilt Initiative in Data Intensive Astrophysics (VIDA) Fellowship. This work was conducted in part using the resources of the Advanced Computing Center for Research and Education (ACCRE) at Vanderbilt University, Nashville, TN. 
\end{acknowledgments}

\appendix

\section{\label{sec:Sampling from an Inverse-Wishart Distribution}Sampling from an Inverse-Wishart distribution}

There are two techniques to sample from a (Inverse) Wishart distribution. The first takes advantage of the definition of a Wishart distribution, in which any positive-definite matrix $\Phi^{-1}$ of size $p \times p$ that follows a Wishart distribution with degrees of freedom $\nu$ and scale-matrix $S^{-1}$ is constructed from the sum of products of column vectors, i.e.,
\begin{equation}
    \Phi^{-1}=\sum\limits_{i=1}^{\nu }{x_{i}{{x}^{T}_{i}}}.
\end{equation}
In the above, each $x_i$ is a column vector with $p$ rows whose underlying distribution is a multivariate normal with zero mean and covariance $S^{-1}$. Hence, to sample from an Inverse-Wishart distribution with degrees of freedom $\nu$ and scale-matrix $S$, one simply constructs $\Phi^{-1}$ using $\{x_i\}$ and inverts the resulting matrix.

The second technique is known as the \emph{Bartlett decomposition} \citep{Bartlett_1934}. To sample from $\Phi^{-1}$, one first samples from a \emph{standard Wishart} distribution, which is a Wishart distribution with degrees of freedom $\nu$ and the identity matrix as its scale-matrix. In \citet{Bartlett_1934}, it is shown that sampling from this standard distribution is equivalent to finding the matrix product of a lower-triangular matrix $A$ with its transpose (i.e., $\Phi^{-1}_{\text{standard}}=AA^T$). The matrix $A$ is constructed as 
\begin{align}
    A=\left[ \begin{matrix}
   \sqrt{\tilde{m}_{\nu }} & 0 & 0 & \cdots  & 0  \\
   {{\tilde{n}}_{21}} & \sqrt{\tilde{m}_{\nu -1}} & 0 & \cdots  & 0  \\
   {{\tilde{n}}_{31}} & {{\tilde{n}}_{32}} & \sqrt{\tilde{m}_{\nu -2}} & \cdots  & 0  \\
   \vdots  & \vdots  & \vdots  & \ddots  & \vdots   \\
   {{\tilde{n}}_{p1}} & {{\tilde{n}}_{p2}} & {{\tilde{n}}_{p3}} & \cdots  & \sqrt{\tilde{m}_{\nu -p+1}}  \\
\end{matrix} \right],
\end{align}
where $\tilde{m}_i$ are samples from a $\chi^2$-distribution with degrees of freedom $i$, and $\tilde{n}_{ij}$ are independent samples from a standard Normal distribution. In order to then convert this draw from a standard Wishart into a Wishart distribution with scale-matrix $S^{-1}$, we perform a square-root factorization of $S^{-1}=LL^T$ such that\footnote{Any similarities of notation with the lead author's surname are entirely coincidental.}
\begin{align}
    \Phi^{-1}=LA{{A}^{T}}L^{T}.
\end{align}
The matrix $\Phi$ is computed by either directly inverting $\Phi^{-1}$ or through inverting the matrix product $LA$. The computational advantage of this remarkable technique is that one deals with square-root factorizations of matrices, rather than the matrices themselves. This ensures numerical stability as well as computational efficiency, since using triangular solvers to find $\Phi^{-1}$ and $\Phi$ by only knowing $A$ and a square-root factorization of $S$ can be performed cheaply and accurately \citep{newwishartsamp}.

\section{\label{sec:math proof}Justifying the choice of prior}

\begin{figure*}[!ht]
    %\centering
\includegraphics[width=\linewidth]{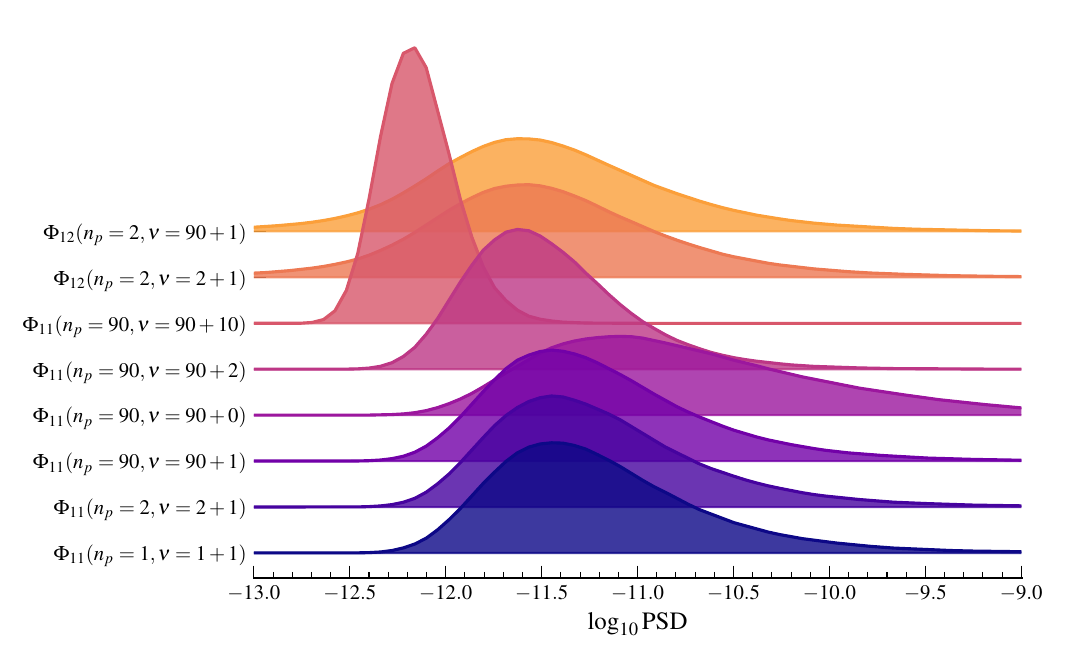}
    \caption{Plots depicting the degree of consistency of conditional posterior probability distributions obtained using Inverse-Wishart distributions with various degrees of freedom $\nu$. The bottom plot is the conditional probability distribution of a single-pulsar PTA. The comparison highlights the bias that one will introduce if $\nu$ is chosen to be different than $n_p+1$. Remarkably, the number of pulsars in the PTA does not affect the posterior-probability so long as $\nu = n_p+1$. This is true even for the cross terms as shown by the top two distributions. Note that to plot $\Phi_{12}$ distributions alongside the rest, we square their values and take the $0.5\log_{10}$ of the result.}
    \label{fig:theo}
\end{figure*}

It is not immediately obvious from \autoref{actual prior} and \autoref{rhogivenb} that the resulting marginalized posterior distributions for $\Phi_{II}$ parameters are invariant under a logarithmic transformation for all $n_p$. Here, we explore this invariance both mathematically and computationally. As the number of frequencies does not make any difference in our discussions in this section, we omit the index $k$ and consider only one frequency.

From \S\ref{sec:Sampling from an Inverse-Wishart Distribution}, we can write
\begin{equation}
   \Phi = {{L}^{-T}}{{A}^{-T}}{{A}^{-1}}{{L}^{-1}},
\end{equation}
such that the diagonal entry for the $I$-th pulsar can be written as
\begin{align}
  {{\Phi }_{II}}&=\sum\limits_{P,R,U=1}^{{{n}_{p}}}{{{\left( {{L}^{-T}} \right)}_{IP}}{{\left( {{A}^{-T}} \right)}_{PR}}{{\left( {{A}^{-1}} \right)}_{RU}}{{\left( {{L}^{-1}} \right)}_{UI}}} \\ 
 &=\sum\limits_{P,R,U=1}^{{{n}_{p}}}{{{\left( {{L}^{-1}} \right)}_{PI}}{{\left( {{L}^{-1}} \right)}_{UI}}{{\left( {{A}^{-1}} \right)}_{RP}}{{\left( {{A}^{-1}} \right)}_{RU}}}.
\end{align}
For the case where we are inspecting the final entry of the diagonal, where $I = n_p$, we have
\begin{align}
   {{\Phi }_{{{n}_{p}}{{n}_{p}}}}&=\sum\limits_{P,R,U=1}^{{{n}_{p}}}{{{\left( {{L}^{-1}} \right)}_{P{{n}_{p}}}}{{\left( {{L}^{-1}} \right)}_{U{{n}_{p}}}}{{\left( {{A}^{-1}} \right)}_{RP}}{{\left( {{A}^{-1}} \right)}_{RU}}} \\ 
 & ={{\left( {{L}^{-1}} \right)}_{{{n}_{p}}{{n}_{p}}}}{{\left( {{L}^{-1}} \right)}_{{{n}_{p}}{{n}_{p}}}}{{\left( {{A}^{-1}} \right)}_{{{n}_{p}}{{n}_{p}}}}{{\left( {{A}^{-1}} \right)}_{{{n}_{p}}{{n}_{p}}}} \\ 
 & = \left[ {{\left( a_{{{n}_{p}}}^c \right)}^{2}}+{{\left( a_{{{n}_{p}}}^s \right)}^{2}} \right]\frac{1}{X_{n_pn_p,\nu-n_p+1}} \label{lastdiag},
\end{align}
where $X_{n_pn_p,\nu-n_p+1}$ is drawn from a $\chi^2$ distribution with $\nu-n_p+1$ degrees of freedom. Note that to arrive at \autoref{lastdiag}, we have used the fact that both $A$ and $L$ are lower-triangular matrices resulting in the non-zero terms to be made by following $I\le P,P\le R,R\ge U,U\ge I$.   
In fact, \autoref{lastdiag} is sufficient to prove that all $\Phi_{II}$ parameters must obey the same distribution as $\Phi$. To prove this point, note that there always exist a \emph{similar matrix} to $\Phi$ through an orthogonal permutation matrix $P$ that permutes the rows and columns of $\Phi$.  This is due to $\Phi$ being a square symmetric matrix. This permutation results in the similar-matrix possessing the same diagonal elements as $\Phi$ but with a, potentially, different order (e.g., $\Phi _{II}^{\text{similar}}={{\Phi }_{JJ}}$ for $I \neq J$). Intuitively, this means that our choice for what we call pulsar $I$ is arbitrary. Thus, all diagonal elements of $\Phi$ must obey the same distribution as $\Phi_{n_pn_p}$ expressed in \autoref{lastdiag}. Lastly, this further implies that by setting $\nu = n_p+1$, one obtains the same posterior-probability distribution as of \autoref{IG}, regardless of $n_p$. Since our choice of prior expressed in \autoref{actual prior} results in the same conditional posterior distribution for all $\Phi_{II}$ as the scale-invariant prior of  
\autoref{sgp} does, we assert that our choice of prior is justified.

To visualize this proof, we fix the value of the Fourier coefficients to
\begin{equation}
    {{a}_{I}}=\sum\limits_{J=1}^{{{n}_{p}}}{{{P}_{IJ}}{{\tilde{n}}_{J}}} \label{theoa},
\end{equation}
where ${{\Phi }^{\text{true}}}={{P}}P^{T}$ (i.e., some fixed red-noise covariance matrix) and $\tilde{n}\sim{\ }\text{Normal}\left( 0,1 \right)$. We then study how the degrees of freedom parameter, $\nu$, of the Inverse-Wishart distribution affects the conditional posterior probability, $p(\Phi|\bm{a})$. This could potentially reveal any inherent biases associated with the specific choice of the prior that we have used in this work.

\autoref{fig:theo} illustrates the impact of the number of pulsars $n_p$ and the degrees of freedom $\nu$ on the conditional posterior probability of $\Phi_{11}$ and $\Phi_{12}$. As expected, the $\Phi_{11}$ conditional posterior probabilities in which $\nu = n_p+1$ are consistent with one other regardless of $n_p$, while other choices with $\nu\neq n_p+1$ are inconsistent. More specifically, the Hellinger distance between the single-pulsar distribution (the lowest distribution on the vertical axis) and those of Inverse-Wishart with $\nu = n_p+1$ is $<0.04$, indicating excellent agreement between the distributions. Remarkably, the cross-terms also obey the same trend when $\nu$ is set to $n_p + 1$, which is evident from the top two distributions in \autoref{fig:theo} for the $\Phi_{12}$ parameter.

\section{\label{sec:GibbsMCMC}Gibbs sampling as a zero-rejection Markov Chain Monte Carlo algorithm}
Our goal in this appendix is to prove that Gibbs sampling with conjugate priors is a special form of MCMC sampling in which the Metropolis-Hastings ratio always evaluates to one, resulting in $100\%$ acceptance of proposed samples.

Consider a model with parameters denoted collectively by $\bm{\theta}$. In Gibbs sampling we split up $\bm{\theta}=\bm{\phi}\cup\bm{\varphi}$, such that a subset of parameters $\bm{\phi}$ is held fixed, while the rest of the parameters $\bm{\varphi}$ is sampled conditioned on the fixed parameters. Let the current value of parameters be $\bm{\theta}^j$ and the next value proposed by the MCMC algorithm be $\theta^{j+1}$.
A typical MCMC algorithm employs the Metropolis-Hastings ratio to determine whether a new sample drawn from a proposal distribution is accepted or rejected as a draw from the target distribution. In our case, this ratio is given by
\begin{equation}
    H = \frac{p\left( \bm{\theta}^{j+1} \right) q\left( \bm{\theta}^j \right)}{p\left( \bm{\theta}^j \right) q\left( \bm{\theta}^{j+1} \right)}, \\ 
\end{equation}
where $p(\bm{\theta}^j)$ is the un-normalized posterior density of $\bm{\theta}^j$, i.e., the product of the likelihood and the prior, ignoring the evidence as a constant for parameter estimation. The factor $q(\bm{\theta}^j)$ is the density of a proposal distribution at $\bm{\theta}^j$; this proposal distribution is typically chosen from a standard family of distributions (and sometimes tuned during sampling) to achieve good acceptance of samples and exploration of the target posterior. Upon evaluating $H$ for a sample, $\bm{\theta}^{j+1}$, drawn from the proposal distribution, the value is then compared to a sample, $u$, drawn from a standard uniform distribution. If $H>u$, $\bm{\theta}^{j+1}$ is accepted as the next point in the MCMC chain; if $H<u$ then the chain remains at $\bm{\theta}^j$, a new point is proposed, and the loop is repeated. 

In Gibbs sampling, random draws from a joint posterior distribution on parameters $\bm{\theta}$ are sampled through sequential updates on constituent (blocks of) parameters $\bm{\varphi}$, conditioned upon the fixed values of the other parameters $\bm{\phi}$. A single iteration of Gibbs sampling would involve a loop through all (groups of) parameters, in which $\bm{\varphi}$ is drawn from $p(\bm{\varphi} | \bm{\phi})$. Despite random samples being drawn from conditional probabilities of each parameter, over many iterations of the Gibbs loop the distribution of samples converges to the target joint posterior. Hence, we now return to the Metropolis-Hastings ratio definition and further split our posterior density factors into conditional posteriors such that 
\begin{align}
    H &=\frac{p(\bm{\varphi}^{j+1},\bm{\phi}^{j+1}) q(\bm{\varphi}^j,\bm{\phi}^j)}{p(\bm{\varphi}^j,\bm{\phi}^j) q( \bm{\varphi}^{j+1},\bm{\phi}^{j+1})} \nonumber\\ 
     & =\frac{p(\bm{\varphi}^{j+1} | \bm{\phi}^{j+1}) p(\bm{\phi}^{j+1})}{p(\bm{\varphi}^j | \bm{\phi}^{j}) p(\bm{\phi}^j)} \frac{q(\bm{\varphi}^j, \bm{\phi}^j)}{q(\bm{\varphi}^{j+1}, \bm{\phi}^{j+1})} \label{Hastings}.
\end{align}
The final step in our proof is to adopt a proposal distribution corresponding to the target posterior itself. We split the proposal density factors such that
\begin{align}
   q(\bm{\varphi}^j,\bm{\phi}^j) &= p(\bm{\varphi}^j | \bm{\phi}^j) p(\bm{\phi}^j), \nonumber \\ 
   q(\bm{\varphi}^{j+1},\bm{\phi}^{j+1}) &= p(\bm{\varphi}^{j+1} | \bm{\phi}^{j+1}) p(\bm{\phi}^{j+1}),
 \end{align}
while $p(\bm{\phi}^{j+1}) = p(\bm{\phi}^j)$ as $\bm{\phi}^{j+1} = \bm{\phi}^j$. The result of this choice of proposal distribution is that the denominator and the numerator of \autoref{Hastings} will cancel each other, leading to a Hastings ratio of $1$, and hence a rate of acceptance of proposed samples equaling $100\%$. In summary, if we are able to propose samples from the conditional posterior of a parameter in a Gibbs loop, then proposal rejection is guaranteed to never happen. This ability to draw samples from conditional posteriors is made possible through conjugate priors, as discussed in the main text.  

\bibliography{main}
\end{document}